\documentclass[10pt, a4paper]{article} 
\usepackage[utf8]{inputenc}
\usepackage[british]{babel}
\usepackage[nottoc]{tocbibind}
\usepackage{authblk}

\usepackage{blindtext}
\usepackage[top=1in, bottom=1in,left=1in,right=1in]{geometry}
\usepackage{amssymb}
\usepackage{amsmath}
\usepackage{mathtools}
\usepackage{multicol}
\usepackage{tikz}

\usepackage{amsfonts}
\usepackage{graphicx}
\usepackage{subfig}
\usepackage{caption}
\usepackage{braket}
\usepackage{array}
\usepackage{float}
\usepackage{hyperref}
\usepackage{cite}
\newcolumntype{M}[1]{>{\centering\arraybackslash}m{#1}}
\newcolumntype{N}{@{}m{0pt}@{}}
\usepackage{enumerate}
\usepackage{bm}
\usepackage{lipsum}

\providecommand{\keywords}[1]
{
  \small	
  \textbf{\textit{Keywords---}} #1
}

\title{The role of thermal and squeezed photons in the entanglement dynamics of the double Jaynes-Cummings model}

\begin{document}

\def\correspondingauthor{\footnote{Corresponding author's email: mandalkoushik1993@gmail.com}}
\author[1]{Koushik Mandal \correspondingauthor{}}
\author[2]{Chandrashekar Radhakrishnan}
\author[1]{M. V. Satyanarayana }
\affil[1 ]{\textit{Department of Physics, Indian Institute of Technology Madras, Chennai, India, 600036}}

\affil[2]{\textit{Department of Computer Science and Engineering, NYU Shanghai, 567 West Yangsi Road, Pudong, Shanghai 200126, China}}
\date{}

\maketitle
\begin{abstract}
The effects of squeezed photons and thermal photons on the entanglement dynamics of atom-atom, atom-field and field-field subsystems are studied for the double Jaynes-Cummings model. For this purpose, squeezed coherent states and Glauber-Lachs states of radiation are chosen as field states. For the atomic states, we choose one of the Bell state as pure state and a Werner-type state as mixed state. Werner-type state is used to understand the effects of mixedness on entanglement. To measure the entanglement between the two atoms, Wootters' concurrence is used; whereas for the atom-field and field-field subsystems, negativity is chosen. The squeezed photons and thermal photons create, destroy and transfer entanglement within various subsystems. Also, the addition of squeezed photons and thermal photons either lengthens or shortens the duration of entanglement sudden deaths (ESD) associated with atom-atom, atom-field and field-field entanglement dynamics in a complementary way. The effects of Ising-type interaction, detuning and Kerr-nonlinearity on the entanglement dynamics are studied. Each of these interactions removes the ESDs associated with various subsystems. We show that new entanglements are created in this atom-field system by introducing Ising-type interaction between the two atoms. With proper choice of the parameters corresponding to Ising-type interaction, detuning and Kerr-nonliearity, entanglement can be transferred among various subsystems.
\end{abstract}


\section{Introduction}
Quantum entanglement is a form of quantum correlation \cite{RevModPhys.81.865} which has become ubiquitous in quantum technologies which are inspired by the second quantum revolution. It has become a resource for quantum information theory \cite{isar1994open, PhysRevA.64.062106, PhysRevA.70.052110, PhysRevLett.99.160502, PhysRevA.83.022109, xu2013experimental, PhysRevB.90.054304, dajka2014disentanglement, Aolita_2015, PhysRevA.92.012315} and several quantum information processing tasks like quantum teleportation \cite{PhysRevLett.70.1895}, super dense coding \cite{PhysRevLett.69.2881}, quantum secret sharing and anonymous transmission \cite{PhysRevLett.67.661}. In most of these applications the entanglement is processed over a finite interval of time and also quantum systems experience decoherence due to the external environment. Under these conditions it becomes necessary to have a complete understanding of the time dynamics of entanglement. In recent years, the study of entanglement dynamics in quantum optics has become a very important topic of research and has drawn much attention in the area of ion traps \cite{PhysRevA.76.041801, PhysRevA.95.043813, PhysRevA.97.043806, PhysRevA.49.1202}, cavity quantum electrodynamics \cite{PhysRevA.82.053832, rosadon, AHMADI2011820, niemczyk2010circuit, haroche2020cavity, fink2008climbing, PhysRevLett.105.173601, PhysRevA.92.023810}, circuit quantum electrodynamics \cite{Mamgain_2023} and linear optical systems. In all these physical devices, the quantum systems are a mix of both atoms and fields. Hence, it is important to study the entanglement dynamics in systems where there is an interaction between atoms and fields. The Jaynes-Cummings model \cite{jaynes1963comparison} is a simple and celebrated scheme, which describes the interaction between an atomic system and a field mode. This model plays a fundamental role in quantum optics and can be realized in practice with Rydberg atoms in high-$Q$ superconducting cavities and trapped ions \cite{nayak1988quantum, gea1990collapse, karatsuba2009resummation, cirac1994quantum, jakubczyk2017quantum, vogel1989k, Chong_2020}. A generalization of the Jaynes-Cummings model known as double Jaynes-Cummings model (DJCM) was done in Ref. \cite{eberly} where the authors consider two non-interacting cavities each containing their own atom and examine the entanglement dynamics between them. There is a sudden loss of total entanglement in the system, a feature which is referred to as entanglement sudden death (ESD)\cite{Yönaç_2006, PhysRevLett.97.140403, eberly2, YU2006393, Yönaç_2007, yu2005evolution, yu2009sudden, eberly3}. Several interesting investigations have been carried out on the DJCM. In particular, in Ref. \cite{li2020entanglement}, the authors study entanglement dynamics of the coherent and squeezed vacuum states and the atoms. It was found that the factors such as atomic spontaneous decay rate, cavity decay rate and detuning have significant effects on the ESD.  Comparison of entanglement dynamics of the standard Jaynes-Cummings model with the intensity-dependent Jaynes-Cummings model was done in Ref. \cite{qin2012entanglement}. To study the effects of interaction, the authors in Ref.  \cite{Pandit_2018,PhysRevA.101.053805} considered cavities with Ising type and photon exchange interactions between the cavities. The notable features of entanglement decay, sudden rebirth and sudden death have been investigated in the presence of intrinsic decoherence have been studied in Ref. \cite{obada2018influence}.  Another work Ref. \cite{laha2023dynamics}, considered a DJCM.  
Here, the author investigates the role of  a beam-splitter, dipole-dipole interaction as well as Ising type interactions.  The photonic modes in this work are considered to be either the photonic vacuum state, or the coherent states or the thermal states and the entanglements investigated are the qubit-qubit and the oscillator-oscillator type. 

In this work, we consider a double Jaynes-Cummings model and investigate the entanglement dynamics of the different possible atom-photon subsystems such as {\it (a)} atom A-atom B, {\it (b)} atom A-field a, {\it (c)} atom A-field  b  and {\it (d)} field a-field b. To study the effects of squeezing and thermal photons on the above mentioned entanglements, the states of radiation field considered are the  squeezed coherent states and Glauber-Lachs states. The squeezed coherent states (SCS)
\cite{ PhysRevA.40.6095,PhysRevA.36.1288,PhysRevA.47.4474, PhysRevA.47.4487, PhysRevA.34.3466, yi1997squeezed, EZAWA1991216, 
PhysRevA.40.2494, RevModPhys.58.1001, PhysRevA.47.5138, PhysRevA.61.010303, PhysRevA.61.022309, PhysRevLett.80.869, PhysRevA.60.937} and the Glauber-Lachs states \cite{PhysRevA.45.5301, sivakumar2012effect} are well known in literature. The initial atomic states investigated are the Bell states (which are pure qubit states) and the Werner type mixed qubit states \cite{PhysRevA.40.4277, czerwinski2021quantifying}. An investigation of entanglement dynamics with atoms in a mixed state in DJCM has not been reported anywhere to the best of our knowledge and through this work we undertake this study. The effects of Ising-type interaction between the atoms, Kerr-nonlinearity and detuning on the entanglement dynamics are also studied. 

The plan of the article is as follows: In Section 2, we introduce the radiation states, atomic states, double Jaynes-Cummings Hamiltonian and the measures of entanglement employed in this work. The entanglement dynamics for the SCS and Glauber-Lachs states of radiation when the atoms are initially in the Bell state and Werner type states are studied in Section 3. Effects of spin-spin Ising interaction on entanglement dynamics is investigated in Section 4. Sections 5 and 6 discuss the effects of detuning and Kerr-nonlinearity on entanglement dynamics respectively. Finally, the conclusions are presented in Section 7.
 
\section{Description of the physical model, the quantum states, and the entanglement measures}

In this section we give an introduction to the DJCM to initiate the discussions into the work. Following this, we list the different radiation states and the atomic states we use in our study. Finally, we give a overview of the different entanglement measures employed in our calculations.  

\subsection{The double Jaynes-Cummings model (DJCM)}
A typical Jaynes-Cummings model consists of a single cavity with a atomic qubit which is interacting with a field mode. While this is a first step towards understanding atom-field interactions, it is a simplistic model without any of the complexities associated with the many-body systems. In order to understand such many-body effects, we need to consider an array of several such cavities each containing an atom interacting with a field mode. While such a model is interesting, an exact analytic calculation would be quite involved. A first step towards understanding interacting cavities is to consider a model in which there are only two cavities each containing an atom interacting with field modes. Such a system is described by the DJCM \cite{eberly} for which the Hamiltonian is given below: 
\begin{align}
\hat{H}_{\text {tot }} = \omega \hat{\sigma}_{z}^{\text{A}}+\omega \hat{\sigma}_{z}^{\text{B}}+ g \left(\hat{a}^{\dagger} 
\hat{\sigma}_{-}^{\text{A}}+\hat{a} \hat{\sigma}_{+}^{\text{A}}\right)+ g \left(\hat{b}^{\dagger} \hat{\sigma}_{-}^{\text{B}}+\hat{b} 
\hat{\sigma}_{+}^{\text{B}}\right)+\nu \hat{a}^{\dagger} \hat{a}+\nu \hat{b}^{\dagger} \hat{b},
\label{djcmmodel}
\end{align}
where A and B are the two atoms in the cavities . The factor $\hat{\sigma}_{z}^{i}$ represents the Pauli matrix in the $z$-basis and
$\hat{\sigma}_{+}^{i}$ and $\hat{\sigma}_{-}^{i}$ are the spin raising and lowering operators respectively. The index $i$ represents the atomic label. The photonic operators $\hat{a}$ and $\hat{b}$ are the annihilation operators corresponding to the two different cavities and the operators $\hat{a}^{\dagger}$ and $\hat{b}^{\dagger}$ are the corresponding creation operators. The coupling constant is represented by $g$ and it describes the strength of the atom-field interaction with $\omega$ and $\nu$ being the atomic transition frequency and the radiation frequency respectively. Let us consider a single cavity Jaynes-Cummings model. The initial state of the atom-field system is considered to be a product state and is of the form $\ket{\psi_{I}(0)} = \ket{e} \otimes \ket{n}$. After a certain period of time the quantum state evolves to 
$\ket{\psi_{I}(t)} = x_{1}(t) \ket{e} \ket{n} + x_{2}(t) \ket{g} \ket{n+1}$, where $x_{1}(t)$ and $x_{2}(t)$ are the probability amplitudes for the system to be found in the excited state $\ket{e} \ket{n}$ and the ground state $\ket{g} \ket{n+1}$ respectively. Solving the Schr\"{o}dinger equation for this state with the initial conditions $x_{1}(0) = 1$ and $x_{2}(0) = 0$, we find  $x_{1}(t) = 
\cos( { g \sqrt{n + 1}\, t})$ and $x_{2} (t) = -i \sin(g \sqrt{n + 1} \, t)$. If the system starts in the initial state 
$\ket{\psi_{I}(0)} = \ket{g} \otimes \ket{n}$, then after the time evolution it becomes $\ket{\psi_{I}(t)} = y_{1}(t) \ket{g} \ket{n} + y_{2}(t) \ket{e} \ket{n-1}$. For the initial conditions we can solve this equation and get $y_{1} (t) = \cos (g \sqrt{n}\, t)$ and $y_{2} (t) =  -i\sin(g \sqrt{n}\, t)$.

In the double Jaynes-Cummings model, the initial state for the atom-field system is 
\begin{eqnarray}
\ket{\psi(0)} = \ket{\psi_{AB}}\otimes \ket{\psi_{F}}_{a}\ket{\psi_{F}}_{b}
= (\cos \alpha \ket{e, g} + \sin\alpha \ket{g, e}) \otimes\left( \sum_{n=0}^{\infty} c_{n}\ket{n} \sum_{m=0}^{\infty} d_{m}\ket{m}\right),
\label{djcm_initialstate}
\end{eqnarray}
where we consider the atomic states corresponding to the cavities a and b are entangled. After time evolution the initial state in 
Eq. (\ref{djcm_initialstate}) evolves to 
\begin{eqnarray}
\ket{\psi(t)} &=& \cos \alpha \sum_{n} c_{n} (x_{1}(t)\ket{e,n} + x_{2}(t) \ket{g, n+1})
\otimes \sum_{m} d_{m} (y_{1}(t)\ket{g,m} + y_{2}(t) \ket{e, m-1}) \\
&& + \sin \alpha \sum_{n} c_{n} (x_{1}(t)\ket{e,n} + x_{2}(t) \ket{g, n+1}) 
\otimes \sum_{m} d_{m} (y_{1}(t)\ket{g,m} + y_{2}(t) \ket{e, m-1}), \nonumber
\end{eqnarray}
which can be rewritten as 
\begin{eqnarray}
\ket{\psi(t)} &=& \sum_{n,m=0}^{\infty} a_{1}(n,m,t)\ket{e, g, n, m} + a_{2}(n,m,t)\ket{e, e, n, m-1} + a_{3}(n,m,t)\ket{g, g, n+1, m} 
                \nonumber \\
              & & + a_{4}(n,m,t)\ket{g, e, n+1, m-1} + a_{5}(n,m,t)\ket{g, e, n, m}  + a_{6}(n,m,t)\ket{g, g, n, m+1} \nonumber\\
              & & + a_{7}(n,m,t)\ket{e, e, n-1, m} + a_{8}(n,m,t)\ket{e, g, n-1, m+1},
\end{eqnarray}
where the factors $a_{i}$'s are as given below: 
\begin{eqnarray}
a_{1}(n,m,t) &=& \cos \alpha \sum_{n} c_{n} x_{1}(t) \sum_{m} d_{m} y_{1}(t); \qquad 
a_{2}(n,m,t) = \cos \alpha \sum_{n} c_{n} x_{2}(t) \sum_{m} d_{m} y_{2}(t);\nonumber \\
a_{3}(n,m,t) &=& \cos \alpha \sum_{n} c_{n} x_{2}(t) \sum_{m} d_{m} y_{1}(t);  \qquad 
a_{4}(n,m,t) = \cos \alpha \sum_{n} c_{n} x_{2}(t) \sum_{m} d_{m} y_{2}(t); \nonumber \\
a_{5}(n,m,t) &=& \sin \alpha \sum_{n} c_{n} y_{1}(t) \sum_{m} d_{m} x_{1}(t);   \qquad 
a_{6}(n,m,t) = \sin \alpha \sum_{n} c_{n} y_{1}(t) \sum_{m} d_{m} x_{2}(t);\nonumber \\
a_{7}(n,m,t) &=& \sin \alpha \sum_{n} c_{n} y_{2}(t) \sum_{m} d_{m} x_{1}(t);   \qquad
a_{8}(n,m,t) = \sin \alpha \sum_{n} c_{n} y_{2}(t) \sum_{m} d_{m} x_{2}(t).\nonumber
\end{eqnarray}
From the knowledge of the time evolved state, one can construct the time evolved density matrix $\rho(t)$ which can be used to calculate the entanglement between different subsystems like the atom-atom, the atom-field and the field-field subsystems.  

\subsection{Photonic States}
In the double Jaynes-Cummings model investigated in our work, we consider two specific forms of photonic states {\it viz} 
{\it (i)} the squeezed coherent states (SCS) and {\it (ii)} Glauber-Lachs states (GLS).  The forms of these states are discussed in this subsection. \\

\noindent {\it Squeezed coherent states: } \\
A single mode continuous variable quantum state can be characterized by two non-commuting observables, whose product of standard deviations obey the uncertainty principle. For such states, the uncertainty takes the form of a circle in the quadrature phase space implying a symmetric distribution of the uncertainties of the observables. By changing the properties of the quantum state it is possible to transform the circular region in the phase space into an elliptical region. This procedure is known as squeezing and when it is carried out on the coherent state we have a squeezed coherent state of the form: 
\begin{equation}
\ket{\alpha,\zeta} = \hat{D}(\alpha)\hat{S}(\zeta)\ket0,
\end{equation}
where the displacement operator $\hat{D}(\alpha)$ and the squeezing operator $\hat{S}(\zeta)$ are 
\begin{equation*}
   \hat{D}(\alpha) = \exp(\alpha\hat{a}^{\dagger} - \alpha^{*}\hat{a}),   \qquad \qquad
   \hat{S}(\zeta) = \exp\left(\frac{1}{2}\zeta\hat{a}^{\dagger2} - \frac{1}{2}\zeta^{*}\hat{a}^{2}\right).
\end{equation*}
The photon creation and annihilation operators are denoted by $\hat{a}$ and $\hat{a}^{\dagger}$. The parameter $\alpha$ being the parameter quantifies the displacement and $\zeta = r \, e^{i\phi}$ , where $r$ is the squeezing parameter. So, the density operator for the SCS is
\begin{equation}
    \hat{\rho}_{\text{scs}} =  \ket{\alpha,\zeta}\bra{\alpha,\zeta}
\end{equation}
and the photon counting distribution is
\begin{equation}
P(n) = \frac{1}{n!\mu}\left(\frac{\nu}{2\mu}\right)^{2} H^{2}_{n}\left(\frac{\beta}{\sqrt{2\mu\nu}}\right) 
       \exp\left(-\beta^{2}\left(1-\frac{\nu}{\mu}\right)\right),
\label{photoncounting}       
\end{equation}
where the parameters described in Eq. (\ref{photoncounting}) are $\mu=\cosh|r|=\sqrt{1+\bar{n}_s}$, $\nu=\sinh|r|=\sqrt{\bar{n}_s}$ 
and $\beta=\sqrt{\bar{n}_c} (\sqrt{1+\bar{n}_s}+\sqrt{\bar{n}_s})$. The quantities $\bar{n}_s$ and $\bar{n}_c$ are the average number of squeezed photons and coherent photons in the system and $H_{n}$ is the Hermite polynomial of order $n$.\\ 

\noindent {\it Glauber-Lachs states: } \\
An interesting form of the radiation mode is the Glauber-Lachs state which is a superposition of the thermal and coherent states. The density operator corresponding to the Glauber-Lachs state is 
\begin{equation}
    \hat{\rho}_\textsubscript{Glauber-Lachs} = \hat{D}(\alpha) \hat{\rho}_\textsubscript{th} \hat{D}^{\dagger}(\alpha),
    \label{Glauber-LachsS}
\end{equation}
where $\hat{D}(\alpha)$ is the displacement operator and $\hat{\rho}_\textsubscript{th}$ is the thermal density operator. The photon-counting distribution for the G-L states given in Ref. \cite{Filipowicz_1986, PhysRevA.45.5301, PhysRevA.35.3433} is 
\begin{eqnarray}
P(n) =\frac{\bar{n}_{th}^{n}}{(1+ \bar{n}_{th})^{n+1}}\exp\left(-\frac{\bar{n}_c}{1+ \bar{n}_{th}}\right) 
       L_{n} \left(-\frac{\bar{n}_c}{\bar{n}_{th}(\bar{n}_{th}+1)}\right),
\end{eqnarray}
where $\bar{n}_{c}$ and $\bar{n}_{th}$ are the average number of coherent photons and thermal photons. The function $L_{n}$ is the Laguerre polynomial of order $n$.

\subsection{Atomic states}
In the double Jaynes-Cummings model, there are two two-level atoms, one in cavity a and the other in  cavity b, respectively. These two-level systems are initially prepared either as a pure state or as a mixed state. For the pure states we consider a maximally entangled Bell state. In our work we use the entangled state of the form
\begin{equation}
\ket{\psi}_\textsubscript{AB} = \cos \theta \ket{e_{\text{A}}, g_{\text{B}}} + \sin\theta \ket{g_{\text{A}}, e_{\text{B}}}
\label{bellstate},
\end{equation}
where $|e_{\text{A}} \rangle$ and $|g_{\text{A}} \rangle$ correspond to the excited and ground states of the atom A. For the mixed states we consider the Werner-type states which are constructed by mixing the maximally entangled states with the maximally mixed states. Considering the Bell state $|\psi^{-} \rangle = \frac{1}{\sqrt{2}} (|g_{\text{A}},e_{\text{B}}\rangle - |g_{\text{B}},e_{\text{A}}\rangle )$, the Werner state\cite{PhysRevA.40.4277}is generated as given below: 
\begin{equation}
    W_\textsubscript{AB} = (1- \lambda)\frac{\textit{I}}{4} + \lambda\, \ket{\psi^{-}}\bra{\psi^{-}}
    \label{WernerBellstate},
\end{equation}
where $\lambda$ is the mixing parameter. When $\lambda = 1$ in Eq. (\ref{WernerBellstate}), the state is the maximally entangled Bell state and for $\lambda =0$ the state is maximally mixed. In the regime $1/3 \leq \lambda \leq 1$, the two qubit Werner state is entangled and for 
$0 \leq \lambda <1/3$ the state is separable. Werner-type mixed states, show non-classical correlations and they can be realized experimentally also by polarization-entangled photon states \cite{PhysRevLett.92.177901}. In ref. \cite{Czerwinski_2021}, the authors have quantified the entanglement for this two-qubit states. Werner states play a crucial role in quantum information theory, quantum teleportation, etc\cite{PhysRevLett.84.4236,PhysRevA.66.062312}. Werner states have also been used in noisy quantum channels\cite{PhysRevA.76.052306}. Therefore, it is important to study the entanglement dynamics of the atom-field system for the Werner states.

\subsection{Entanglement measures}
To characterize the dynamics of entanglement, we need to measure the entanglement in the system. In our work we investigate the dynamics of the bipartite entanglements like the atom-atom entanglement, atom-field entanglement and the field-field entanglement. The atom-atom entanglement can be conclusively measured using concurrence defined in \cite{wootters2001entanglement}
\begin{equation}
C_{\text{AB}} = \text{max}\{0, \xi_{1} - \xi_{2}-\xi_{3}-\xi_{4}\},
\end{equation}
where $\xi_{i} (i = 1, 2, 3,4)$ are the decreasingly ordered square roots of the eigenvalues of the matrix 
$\hat{\rho} \left(\hat{\sigma}_{y}^{\text{A}}\otimes \hat{\sigma}_{y}^{\text{B}}\right) \hat{\rho}^{*}\\ 
\left(\hat{\sigma}_{y}^{\text{A}} \otimes \hat{\sigma}_{y}^{\text{B}}\right)$ and $\hat{\rho}$ is the two qubit atom-atom reduced density matrix. The value of concurrence lies in the range $0 \leq C \leq 1$, where $C=0$ implies a separable state and $C=1$ denotes a maximally entangled state. Though concurrence can be used to compute entanglement in both pure and mixed states, it works only for $2 \otimes 2$ systems, so for higher dimensional systems we need to use other measures. In particular, when we consider the atom-field and field-field subsystems we are looking at $2 \otimes \infty$ and bipartite continuous variable systems. For these systems it is convenient to use the negativity \cite{wei2003maximal} which is defined as 
\begin{equation}
N(t)=\sum_{k}\Big( |\xi_{k}|-\xi_{k} \Big)/2, 
\end{equation}
where $\xi_{k}$ are the eigenvalues of $\hat{\rho}^{\text{PT}}$, the partial transpose of the density matrix, i.e., the matrix which is 
transposed with respect to any one of the subsystems.

\section{Entanglement dynamics in the non-interacting double Jaynes-Cummings model}

\begin{figure}
    \centering
    \includegraphics[scale = 0.4]{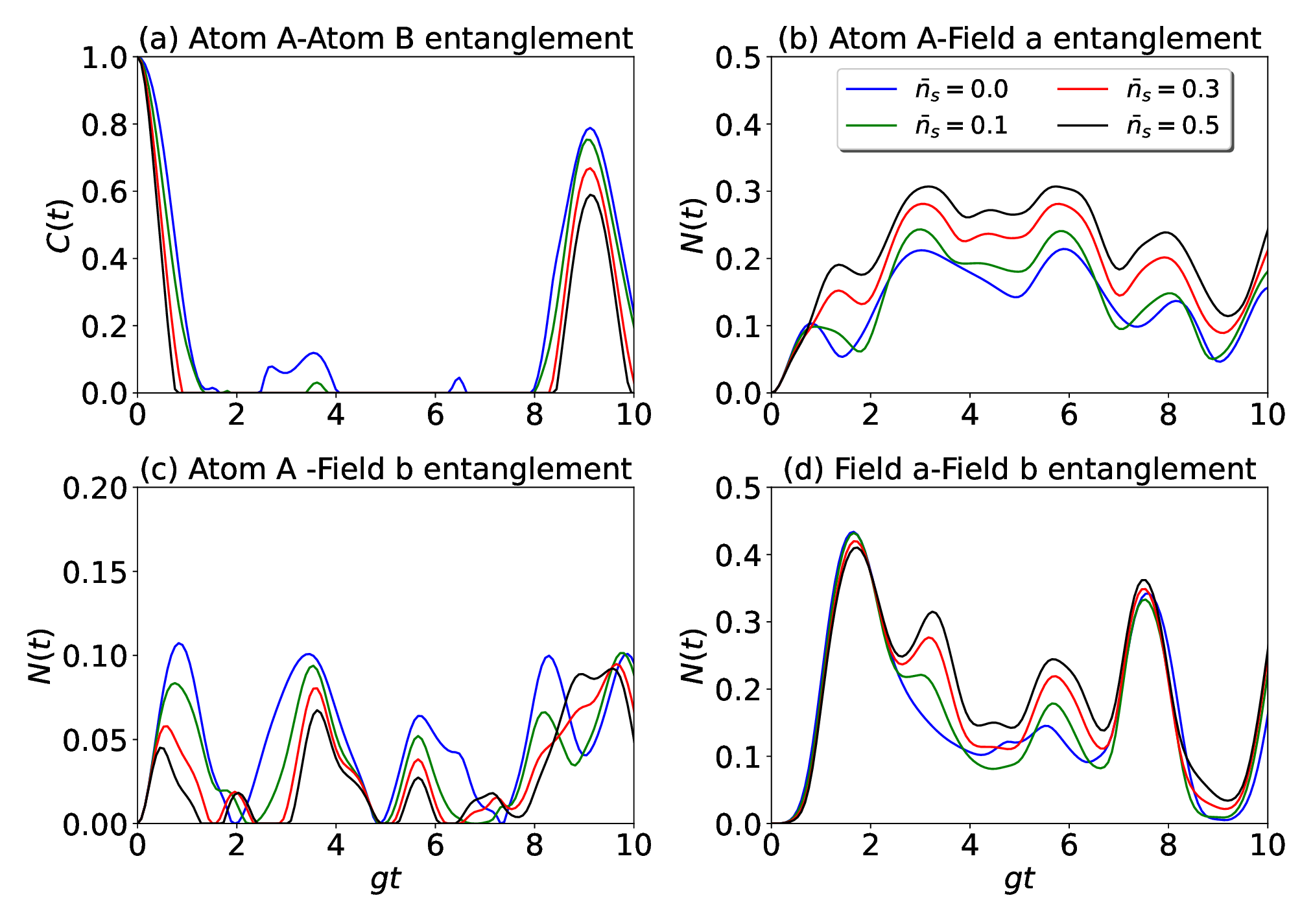}
    \caption*{\textbf{Figure 1.} Entanglement dynamics of atom A-atom B, atom A-field a, atom A-field b and field a-field b subsystems in DJCM with atomic states in the Bell state $\ket{\psi}_\textsubscript{AB}$ and field state in SCS. The other parameters are $\bar{n}_{c} = 0.5$, $\theta = \frac{\pi}{4}$ and $\bar{n}_{s} = 0.0, 0.1, 0.3, 0.5$.}
\end{figure}

\begin{figure}
    \centering
    \includegraphics[scale = 0.45]{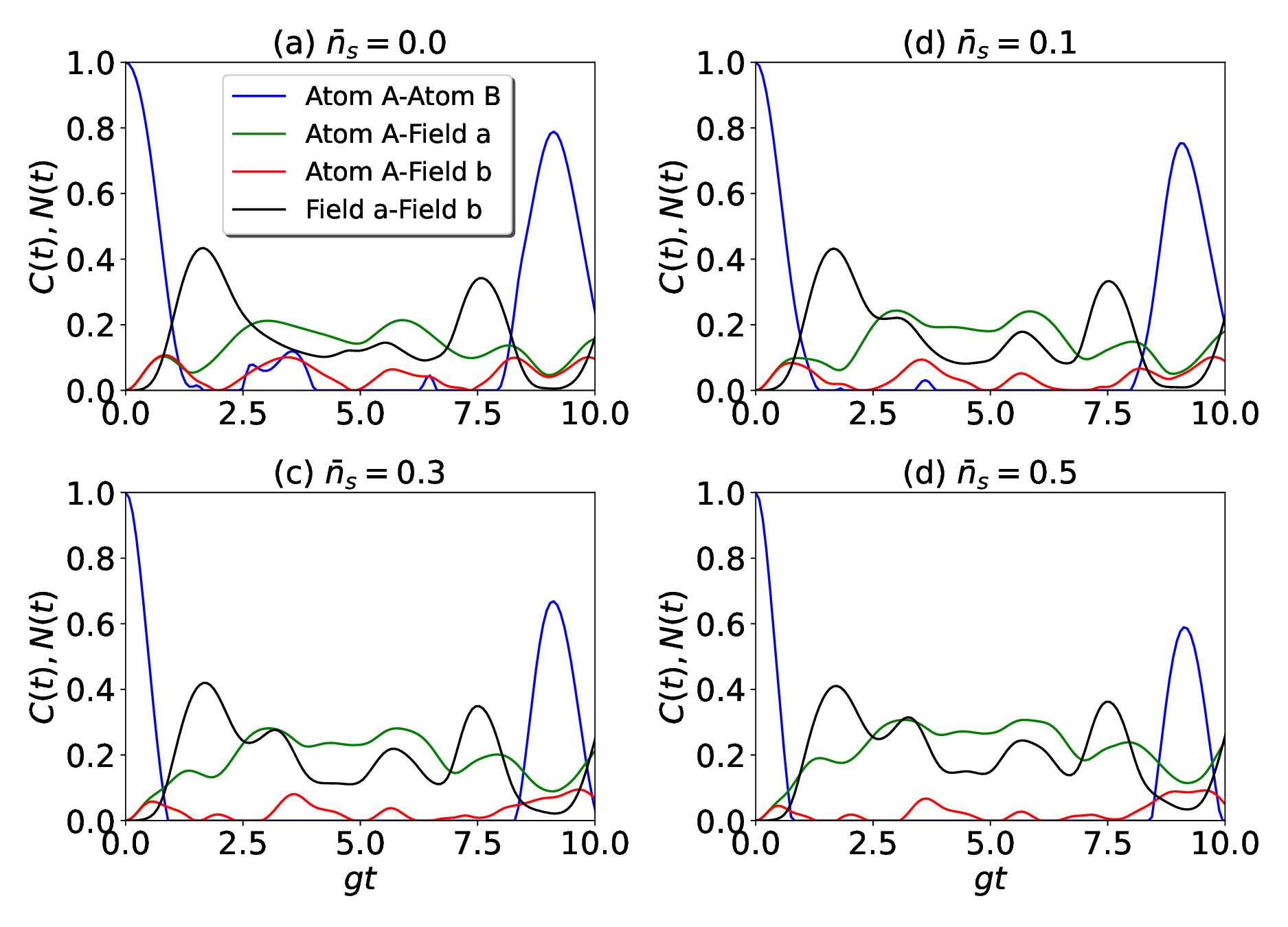}
    \caption*{\textbf{Figure 2.} In this figure, entanglements for all the subsystems are plotted in a single plot for various $\bar{n}_{s}$ in DJCM with atomic states in the Bell state $\ket{\psi}_\textsubscript{AB}$ and field state in SCS. }
\end{figure}

In the DJCM discussed in the  previous section, we consider entangled atomic states which are interacting with different states of the radiation field. Due to this interaction, the entangled states evolve with time and in this section we characterize the temporal evolution of the entanglement.  

\subsection{Squeezed Coherent States}

\subsubsection{For Bell states}
The entanglement dynamics of the double Jaynes-Cummings models with non-interacting cavities is described in Figs. 1 and 2, for the squeezed coherent states. In this subsection we consider the initial state to be a pure state of the form given in Eq. (\ref{bellstate}) which is a Bell state. For the plots we fix the average number of coherent photons ($ \bar{n}_c = 0.5$) and study the dynamics at different average values of squeezed photons at $\bar{n}_s = 0.0, 0.1, 0.3$ and $0.5$. There are four possible bipartite entangled combinations, {\it viz} {\it (a)} atom A-atom B entanglement, {\it (b)} atom A-field a entanglement {\it (c)} atom A-field b entanglement and {\it (d)} field a-field b entanglement.

In Fig. 1 we plot the four different entanglements as subplots from $1$(a) to $1$(d) respectively. From Fig. 1(a) we observe that the atom-atom entanglement falls very sharply with time for all $\bar{n}_{s}$ and consequently ESD appears in the dynamics. This sudden death feature is followed by an entanglement revival. On increasing $\bar{n}_{s}$, the time gap between the ESD and revival increases. Hence, the largest duration for ESD is seen for $\bar{n}_{s} = 0.5$. Also, the height of the revival peak decreases with increase in $\bar{n}_{s}$. The atom A-field a entanglement is given through Fig. 1(b). Here, we notice that the initial entanglement is zero and the entanglement increases with time and also there is no observation of ESD. Also, the amplitude of entanglement increases with $\bar{n}_{s}$. The dynamics of entanglement between the atom A and field b is presented in Fig. 1(c), where we also observe ESD. The frequency of collapse and revival of entanglement is larger compared to the atom A-atom B entanglement. Finally, we plot the entanglement between the field a and field b in Fig. 1(d). The transient dynamics of entanglement between the fields is complementary to the entanglement dynamics between the atoms. When the atom A-atom B entanglement decreases, the field a-field b entanglement increases and vice versa.  

A comparative study of all the four different entanglements for a fixed $\bar{n}_{s}$ is given in Fig. 2, where the subplots Fig. 2(a) to 2(d) depict the entanglement dynamics for $\bar{n}_{s}=0$, $\bar{n}_{s}=0.1$, $\bar{n}_{s}=0.3$ and $\bar{n}_{s}=0.5$ respectively. We can see that at $t=0$, only the atom A-atom B entanglement is present, but as time progresses the other entanglements emerge at the cost of the atom-atom entanglement, which decreases. In fact, the complementarity between the atom-atom and field-field entanglement is very clearly seen in Fig. (2), where we find that the points at which the atom-atom entanglement low are same as the points where the field-field entanglement is high. 

So, from this analysis it can be concluded that the addition of squeezing in the system creates entanglement in atom A-field a, field a-field b subsystems and also entanglement gets transferred from atom-atom subsystem to other subsystems. In Ref.\cite{li2020entanglement} and in Ref. \cite{laha2023dynamics}, it is observed that if the field is in a coherent state, every subsystems suffers ESDs and duration of these ESDs get increased with increasing coherent photons. While, in these cases, entanglement get transferred from one subsystem to another, however, increase in coherent photons does not create new entanglement.

\begin{figure}
    \centering
    \includegraphics[scale = 0.4]{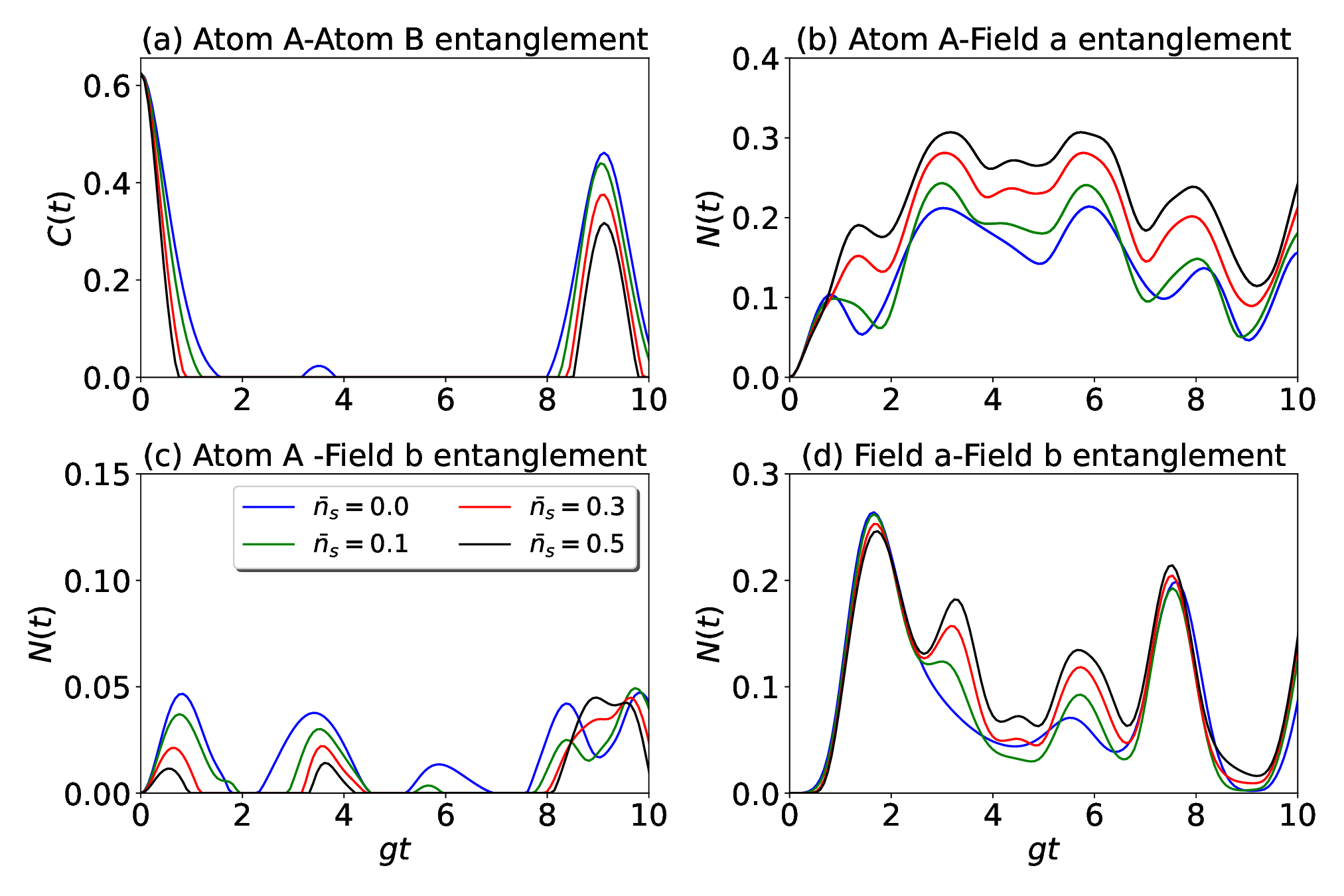}
    \caption*{\textbf{Figure 3.} Entanglement dynamics of atom A-atom B, atom A-field a, atom A-field b and field a-field b subsystems in DJCM with atomic states in Werner state and field state in SCS. The other parameters $\bar{n}_{c} = 0.5$, $\lambda = 0.75$ and $\bar{n}_{s} = 0.0, 0.1, 0.3, 0.5$.}
\end{figure}

\begin{figure}
    \centering
    \includegraphics[scale = 0.4]{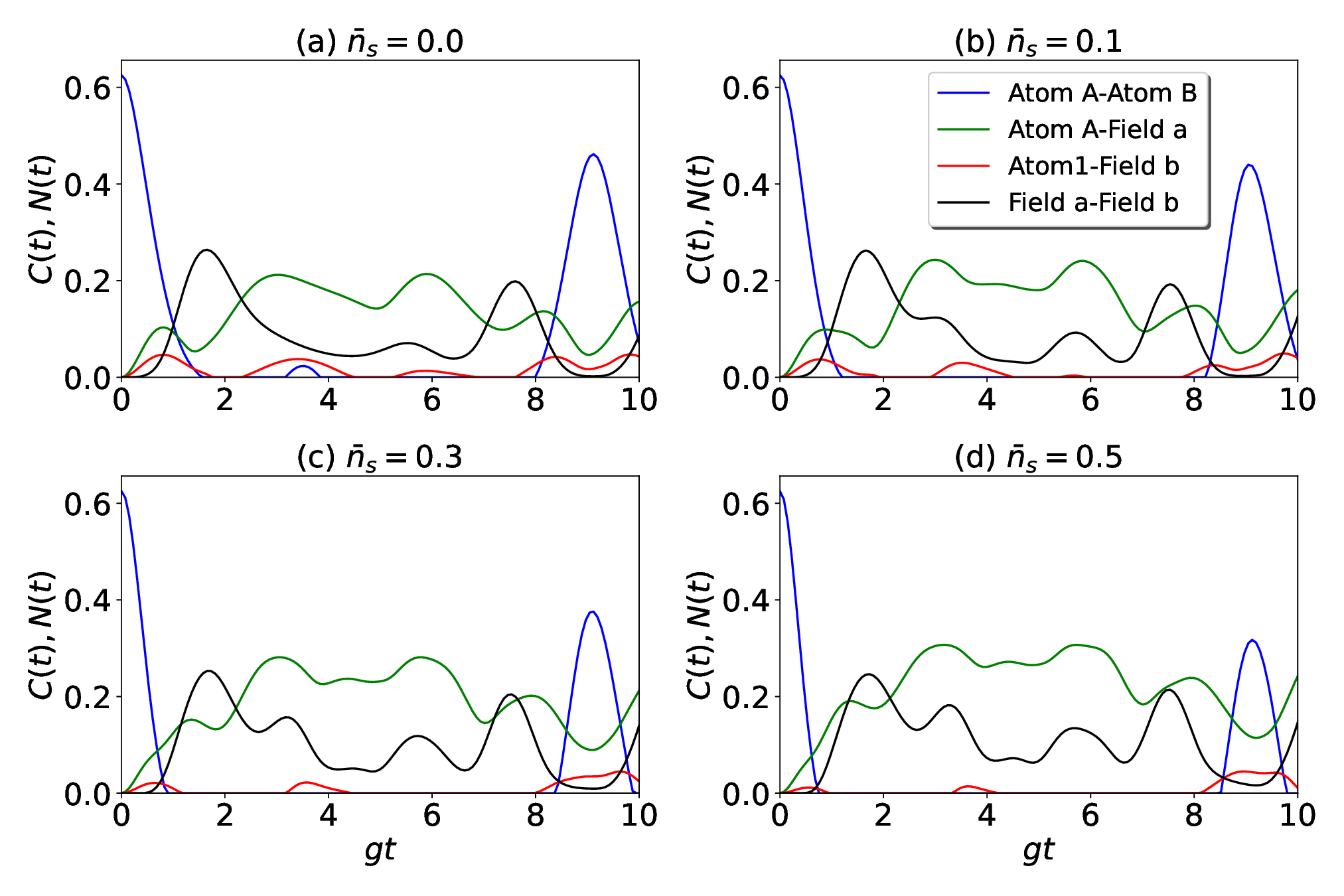}
    \caption*{\textbf{Figure 4.} In this figure, entanglements for all the subsystems are plotted in a single plot for various $\bar{n}_{s}$ in DJCM with atomic states in Werner state and field state in SCS.}
\end{figure}

\begin{figure}
    \centering
    \includegraphics[scale = 1]{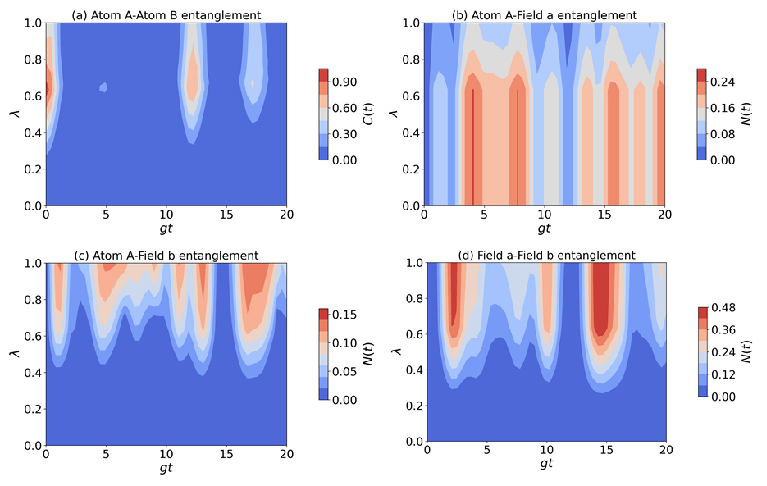}
    \caption*{\textbf{Figure 5.} Contour plots of entanglement dynamics $C(t)$ and $N(t)$ $vs$ $gt$ and $\lambda$ of different subsystems for SCS with atoms in Werner state. Values of the parameters used in these plots are $\bar{n}_c = 0.5, \bar{n}_{s} = 0.1$.}
\end{figure}

\subsubsection{For Werner state}
To investigate the entanglement dynamics of mixed states, we consider the atom A and atom B to form a Werner-type state Eq.(\ref{WernerBellstate})) and the field inside both the cavities are in squeezed coherent states.  

In Fig. 3, the four different bipartite entanglements, namely the atom A-atom B entanglement, the atom A-field a entanglement, the atom A-field b entanglement and the field a-field b entanglement are displayed in the subplots Fig. 3(a)-(d) respectively for a Werner mixing parameter $\lambda = 0.75$. The mixed state dynamics shows features which are very similar to the pure state dynamics. The atom-atom entanglement dynamics shows ESD as well as entanglement revival. While the atom A-field a entanglement does not exhibit sudden death and revival, the entanglement dynamics of atom A-field b displays these interesting features of death and revival. Finally we see that the field a-field b entanglement is initially zero which then increases and then oscillates.  

For four values of average number of squeezed photons $\bar{n}_{s}=0$, $\bar{n}_{s}=0.1$, $\bar{n}_{s}=0.3$ and $\bar{n}_{s}=0.5$ the entanglement plots are given in Fig. 4(a) to 4(d) respectively. In these plots we comparatively analyze the four different bipartite entanglements. It is to be noted that at the points where the atom-atom entanglement reaches a minimum or disappears, the field-field entanglement is maximum and vice versa.  This complementarity suggests that the entanglement which is present in the initial atom-atom bipartite subsystem flows to other bipartite subsystems.  

For further understanding of the entanglement flow, the dynamics of entanglement is investigated, for Werner state parameter in the range $0\leq \lambda \leq 1$. The results are presented in Fig. 5 (a)-(d) which are contour plots of the entanglement as a function of the mixing parameter $\lambda$ and time $t$. The Werner states for $\lambda < 1/3$ are not entangled and this is evident from the plot Fig. 5(a) - that the entanglement vanishes for the atom-atom state. Also, the entanglement shows sudden death, and the revival features for nearly all values of $\lambda$, for which the entanglement is present. When we observe the entanglement dynamics of atom A-field a bipartite system, we find a finite amount of entanglement for all possible values of $\lambda < 1/3$ which signals the generation of entanglement due to time dynamics. Finally when we look at the remaining bipartite combinations of atom A-field b and field a-field b, we find that entanglement vanishes for $\lambda < 1/3$. This is because in these bipartite combinations the two members are not directly interacting with each other. The sudden death features are also present in the atom A-field b combination and the field a-field b combination for all values of $\lambda$ where entanglement is present. 

\begin{figure}
    \centering
    \includegraphics[ scale = 0.4]{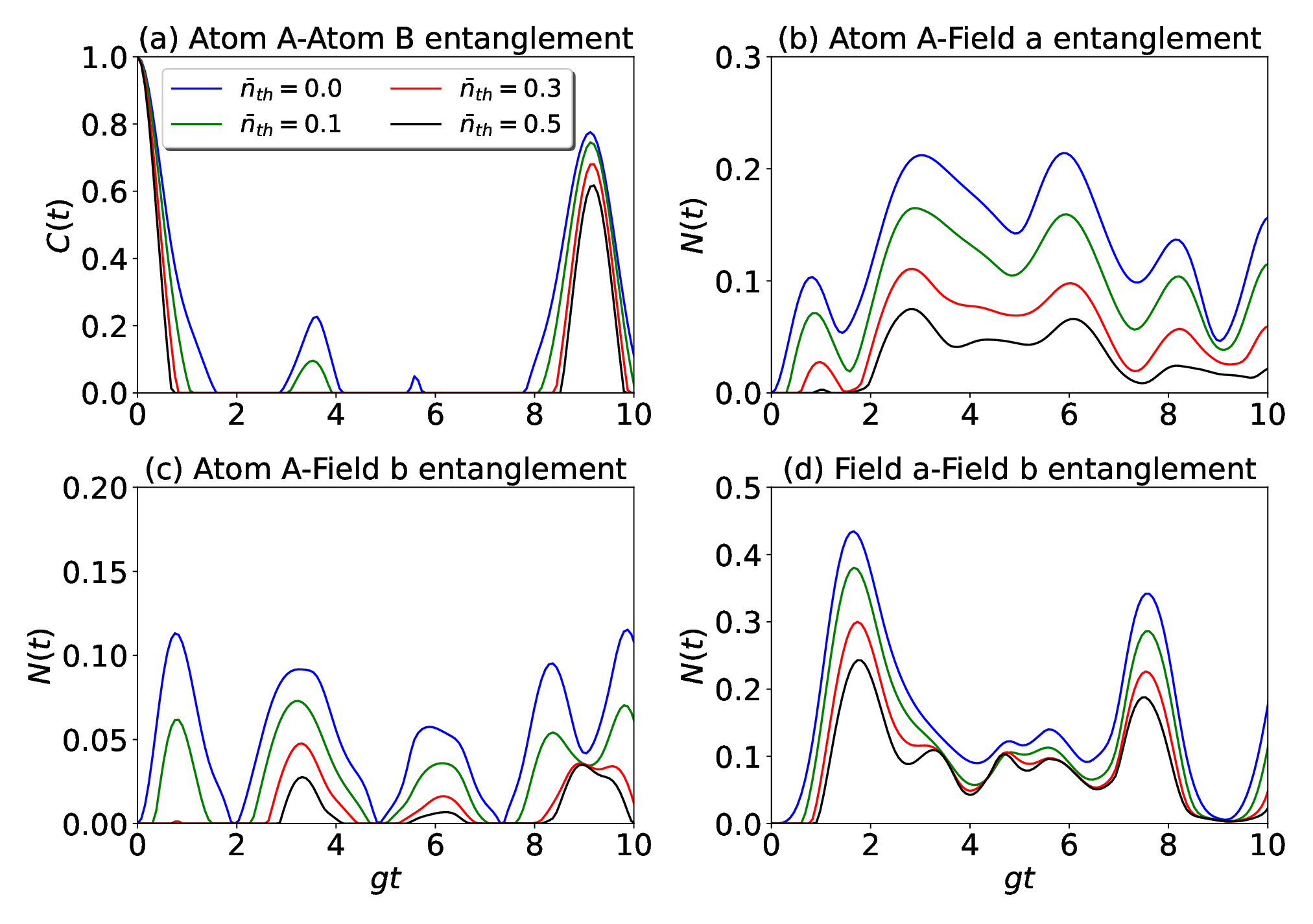}
    \caption*{\textbf{Figure 6.} Entanglement dynamics of atom A-atom B, atom A-field a, atom A-field b and field a-field b subsystems in DJCM with atomic states in the Bell state $\ket{\psi}_\textsubscript{AB}$ and field state in Glauber-Lachs states. Other parameters are $\bar{n}_{c} = 0.5$, $\theta = \frac{\pi}{4}$ and $\bar{n}_{th} = 0.0, 0.1, 0.3, 0.5$.}
\end{figure}

\begin{figure}
    \centering
    \includegraphics[scale = 0.45]{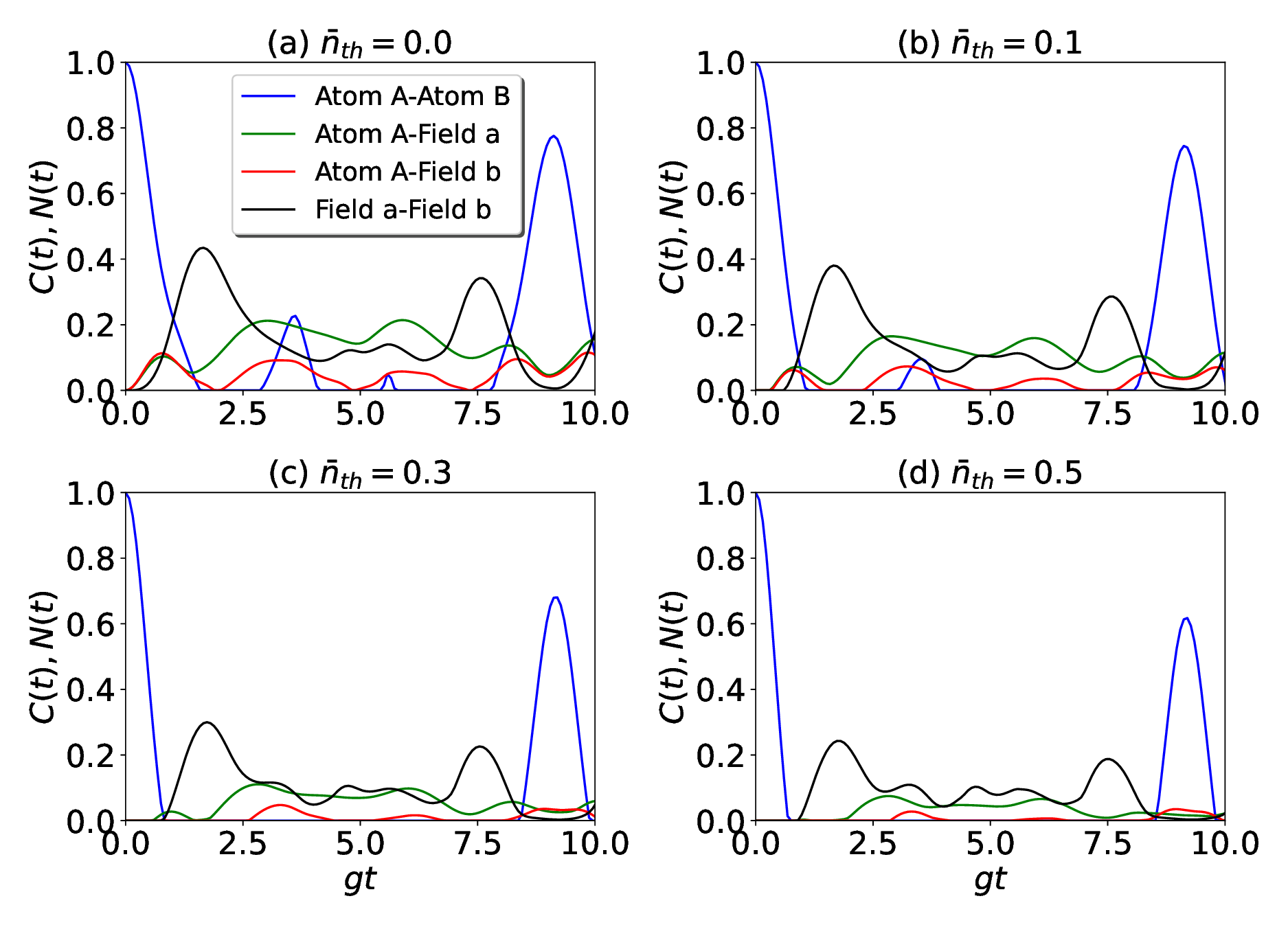}
    \caption*{\textbf{Figure 7.} Here, entanglements for all the subsystems are plotted in a single plot for various $\bar{n}_{th}$ in DJCM with atomic states in the Bell state $\ket{\psi}_\textsubscript{AB}$ and field state in Glauber-Lachs states.}
\end{figure}

\subsection{For Glauber-Lachs states}
The transient dynamics of entanglement of the Glauber-Lachs state is discussed in this subsection. The results corresponding to the atomic pure and mixed states are discussed in two different subsections.  

\subsubsection{For Bell state}
The time evolution of entanglement when the initial states are maximally entangled pure states is given in Fig. 6 for the four possible bipartite systems namely the atom A-atom B entanglement, atom A-field a, atom A-field b and field a-field b entanglements are given through the subplots Fig. 6(a)-(d) respectively. Here, the average number of coherent photons $ \bar{n}_c = 0.5$ and in each plot $\bar{n}_{th}$ is varied as $\bar{n}_{th} = 0.0, 0.1, 0.3$ and $0.5$.  In the plot Fig. 6(a) we find that the entanglement evolution displays sudden death and revival for the atom A-atom B bipartite system. Similar effect is observed for the atom A-field b bipartite system (Fig. 6(c)), but the atom A-field a bipartite system and the field a-field b subsystem in Figs. 6(b) and 6(d) respectively, do not exhibit ESD and revival. The entanglement decreases with increase in  $\bar{n}_{th}$, since the thermal photons cause decoherence in the system.  

In Fig. 7(a)-7(d), we plot the four bipartite entanglements for different values of $\bar{n}_{th}$. A complimentary behavior is observed between the atom A-atom B and the field a-field b entanglement evolution. Also, from the four different bipartite entanglement evolutions, we observe that the initial entanglement between the atoms leads to creation of entanglement between the other bipartite combinations.

\begin{figure}
    \centering
    \includegraphics[scale = 0.4]{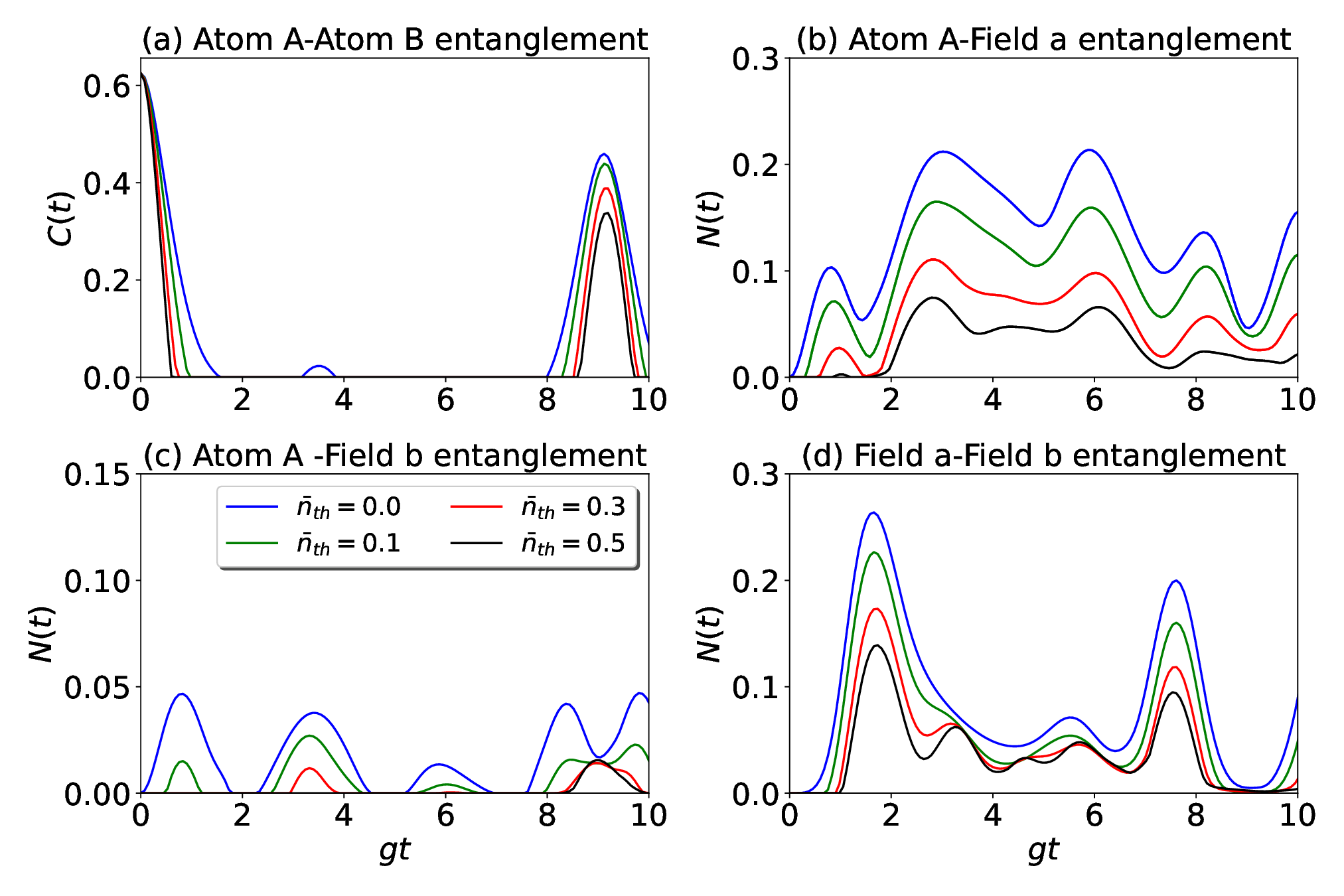}
    \caption*{\textbf{Figure 8.} Entanglement dynamics of atom A-atom B, atom A-field a, atom A-field b and field a-field b subsystems in DJCM with atomic states in Werner state and field state in Glauber-Lachs states. The other parameters $\bar{n}_{c} = 0.5$, $\lambda = 0.75$ and $\bar{n}_{th} = 0.0, 0.1, 0.3, 0.5$.}
\end{figure}

\begin{figure}
    \centering
    \includegraphics[scale = 0.4]{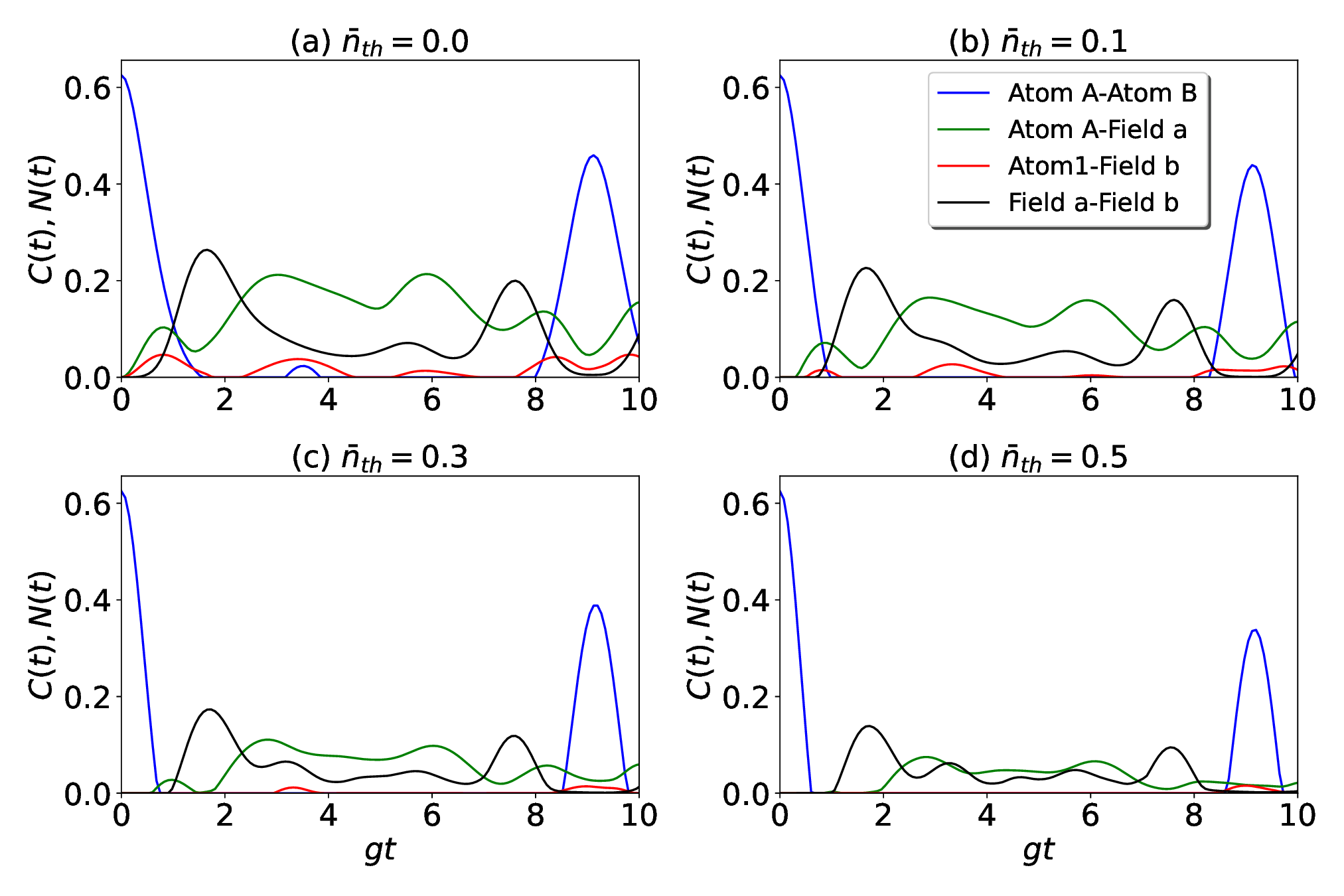}
    \caption*{\textbf{Figure 9.} In this figure, entanglements for all the subsystems are plotted in a single plot for various in DJCM with atomic states in Werner state and field state in Glauber-Lachs states.}
\end{figure}

\begin{figure}
    \centering
    \includegraphics[scale = 1.0]{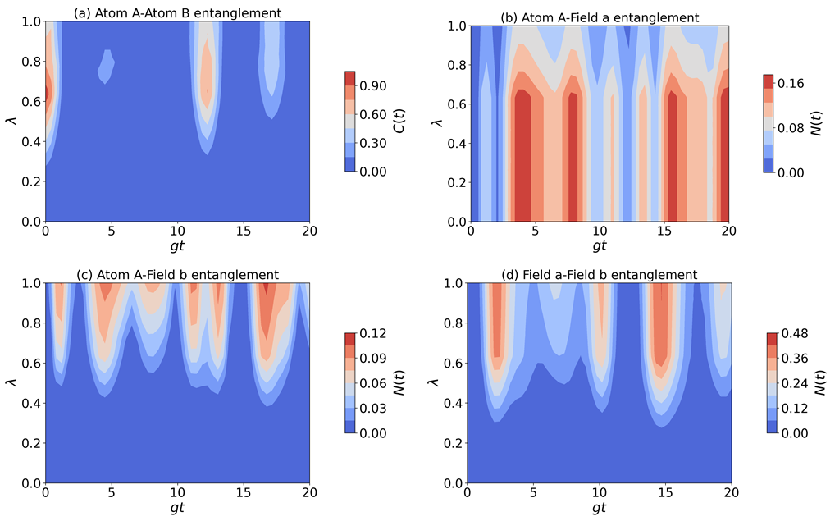}
    \caption*{\textbf{Figure 10.} Contour plots of entanglement dynamics $C(t)$ and $N(t)$ $vs$ $gt$ and $\lambda$ for Glauber-Lachs states with the atoms in Werner state. Values of the parameters used in the plot are $\bar{n}_c = 0.5, \bar{n}_{th} = 0.1$.}
\end{figure}

\subsubsection{For Werner state}
When the initial atom-atom entangled state is a mixed state of Werner type and the field  under consideration are the Glauber-Lachs radiation modes, the dynamics of entanglement is given in Figs. 8, 9 and 10 respectively.  For our investigation, we consider $ \lambda = 0.75$  in Figs. 8 and 9. In Fig. 8(a)-(d), we plot the atom A-atom B entanglement, the atom A-field a entanglement, the atom A-field b entanglement and the field a-field b entanglement respectively. The results are very identical to those of the pure states. The salient features in the pure state case, namely the ESD and revival in the atom A-atom B and atom A-field b as well as the complimentarity in the evolution between the atom A-atom B entanglement and the field a-field b entanglement are well marked out in the entanglement dynamics. The comparative analysis of the entanglement dynamics is given in Fig. 9 (a)-(d) for different values of $n_{th} = 0, 0.1, 0.3$ and $0.5$ respectively which again confirms that the initial entanglement in the atomic pair leads to the creation of entanglement between the other bipartite combinations.  

The variation of entanglement with time as well as the mixing parameter $\lambda$ is given in the contour plot in Fig. 10. In the case of the atom A-atom B entanglement in Fig. 10 (a), it is observed that the initial entanglement is present for all values of $\lambda > 1/3$. The sudden death and revival of entanglement gives rise to islands of entangled regions in the contour plot. In the entanglement dynamics of atom A-field a bipartite system, we find entanglement for all possible values of $\lambda$. For the bipartite combinations {\it viz} the atom A-field b and field a and field b system, it is noticed that the entanglement vanishes for $\lambda < 1/3$. This is due to the fact that the initial atomic systems are no longer entangled and hence they are not capable of creating entanglement in these subsystems. Further, the ESD and revival are present in atom A-atom B as well as the atom A-field a and atom A-field b bipartite systems. 

\section{Entanglement dynamics with Ising interactions}
The investigations carried out in section 3 was on DJCM with no interaction between the two atoms. The Ising type interaction between the two atoms is introduced in this section and its effects on the entanglement dynamics are studied here. The DJCM with Ising type interaction is given by
\begin{equation}
   \hat{H}_{\text{tot}} = \omega \hat{\sigma}_{z}^{\text{A}} + \omega \hat{\sigma}_{z}^{\text{B}} 
 + g (\hat{a}^{\dagger} \hat{\sigma}_{-}^{\text{A}} + \hat{a} \hat{\sigma}_{+}^{\text{A}}) 
+ g (\hat{b}^{\dagger} \hat{\sigma}_{-}^{\text{B}} + \hat{b} \hat{\sigma}_{+}^{\text{B}})
+ \nu \hat{a}^{\dagger} \hat{a} + \nu \hat{b}^{\dagger} \hat{b}
+ J_{z}\hat{\sigma}_{\text{A}}^{z} \otimes \hat{\sigma}_{\text{B}}^{z},
\end{equation}
where $\omega$ and $\nu$ are the atomic and photonic frequencies respectively and $g$ is the atom-photon coupling strength. The factor $\hat{\sigma}_{i}^{z}$ is the Pauli matrix in the $z$-basis and $\hat{\sigma}_{i}^{+}$ and $\hat{\sigma}_{i}^{-}$ are the raising and lowering operator of the atomic system where $i = $(A,B) distinguishes the two atoms. The operators $(\hat{a}^{\dagger},\hat{a})$ and 
$(\hat{b}^{\dagger},\hat{b})$ are the creation and annihilation operators of the photonic systems in cavities a and b respectively. The final term in the Hamiltonian is the Ising interaction term in which $J_z$ is the coupling strength between the two atoms and has the units of energy. The entanglement dynamics of this model is studied for two different types of photonic systems namely, the squeezed coherent state and the Glauber-Lachs state.  

\begin{figure}
    \centering
    \includegraphics[scale = 0.4]{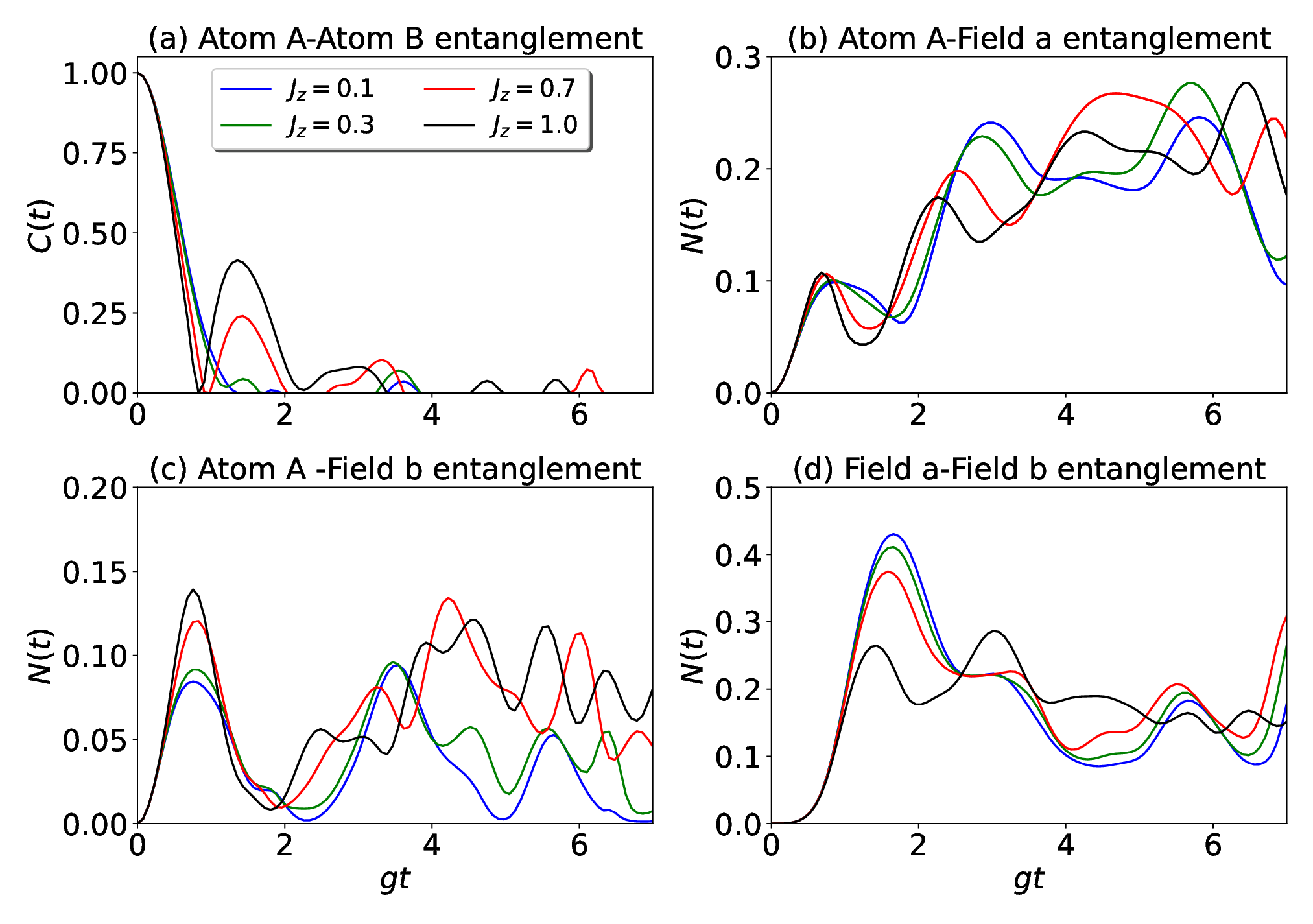}
    \caption*{\textbf{Figure 11.} Effects of spin-spin Ising interaction entanglement dynamics of atom A-atom B, atom A-field a, atom A-field b and field a-field b in DJCM for atomic states in the Bell state $\ket{\psi}_\textsubscript{AB}$ and field state in SCS. The other parameters are $\bar{n}_{c} = 0.5,\bar{n}_{s} = 0.1$, $\theta = \frac{\pi}{4}$ and $J_{z} = 0.1, 0.3, 0.7, 1.0$.}
\end{figure}
The entanglement dynamics of DJCM with Ising-type interaction when the field mode is either a coherent state or a thermal state is investigated in Ref. \cite{laha2023dynamics}. In our work, we consider a field mode which is either a squeezed coherent state or a Glauber Lachs state and do not use any approximation which consequently captures the entire dynamics of the entanglement. 

\subsection{Squeezed Coherent States}
The entanglement dynamics of the interacting double Jaynes-Cummings models where the radiation mode is a squeezed coherent state is discussed in this section and the results corresponding to the Bell state and Werner state are given as two separate subsections. 

\subsubsection{For Bell state}
The time evolution of entanglement which is initially present between the two atoms is described here for an initial state as  a Bell state of the form given in Eq. \ref{bellstate}. The results are described through the plots in Fig. 11, in which the subplot, 11(a) depicts the atom A-atom B entanglement, with  11(b) and 11(c) describing the atom A-field a and atom A-field b entanglement and finally, Fig. 11(d) sketches the field a-field b entanglement.  

We set $\bar{n}_c = 0.5$, $\bar{n}_s = 0.1$ and compare the dynamics for different interaction values namely $J_{z} = 0.1, 0.3, 0.7, 1.0$ in Fig. 11.  While the ESD and entanglement revival appear in the system, the time duration for which the entanglement disappears has decreased. It can be seen that the entanglement between the atoms is preserved for a longer duration, indicating that the Ising type interaction has a positive influence on the entanglement. From Fig. 11(b), it is observed that the atom A-field a entanglement does not change much with the addition of Ising type interaction.  Meanwhile, from Fig. 11(c), we notice that the ESD is no longer present. Also, the amplitude of the entanglement depends on the strength of the interaction. Finally, it is noticed that the field a-field b entanglement which is initially zero increases to a finite value. The amplitude of the entanglement decreases with increase in the strength of interaction. Thus, it is observed that the number of ESDs are much less in the presence of Ising interaction. So, the Ising interaction tries to preserve the entanglement between the atoms.

\begin{figure}
    \centering
    \includegraphics[scale = 0.4]{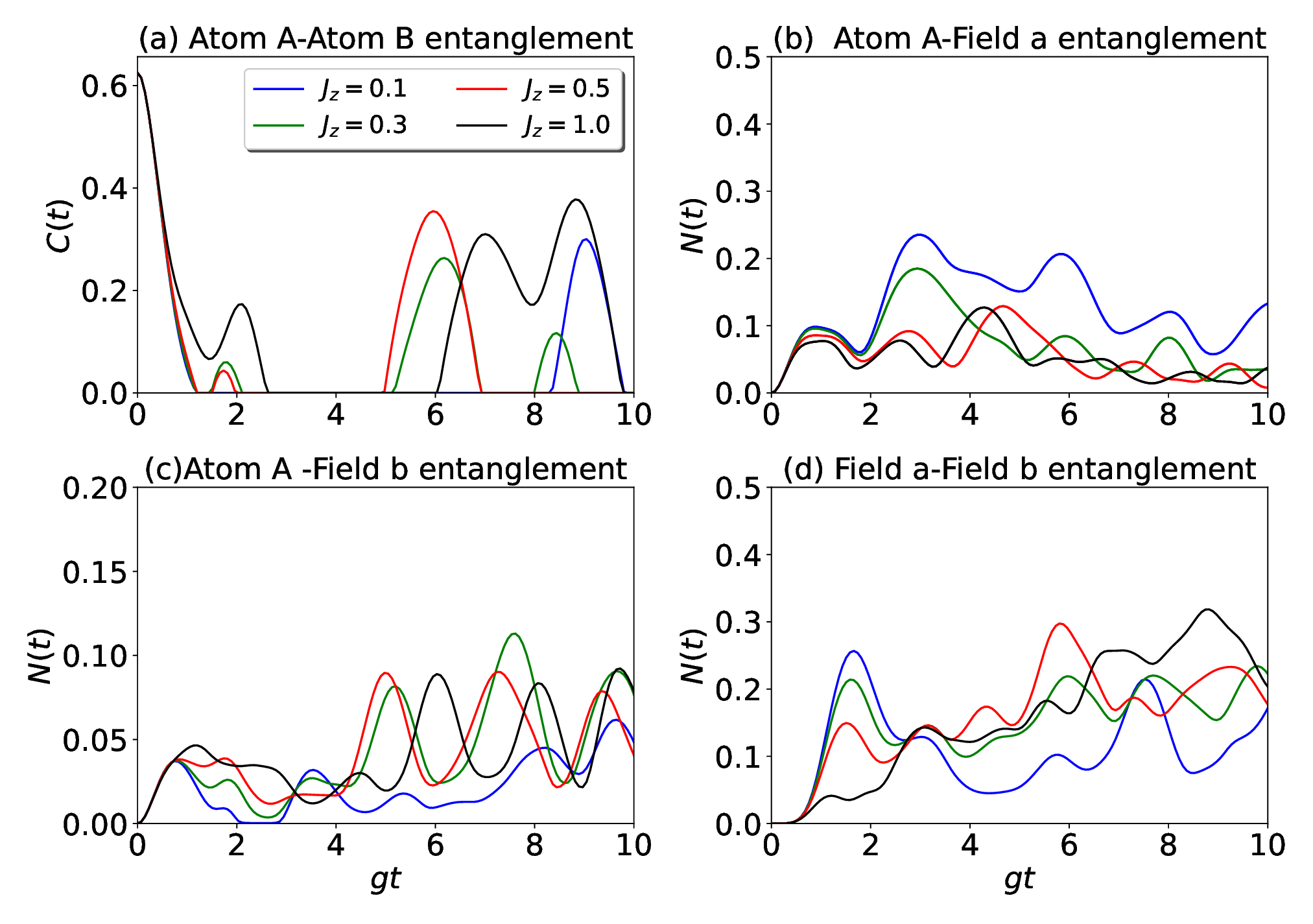}
    \caption*{\textbf{Figure 12.} Effects of spin-spin Ising interaction on entanglement dynamics of atom A-atom B, atom A-field a, atom A-field b and field a-field b subsystems in DJCM with atomic states in Werner state and field state in SCS. Other parameters are $\bar{n}_{c} = 0.5,\bar{n}_{s} = 0.1$, $\lambda = 0.75$ and $J_{z} = 0.1, 0.3, 0.7, 1.0$.}
\end{figure}

\begin{figure}
    \centering
    \includegraphics[scale = 0.5]{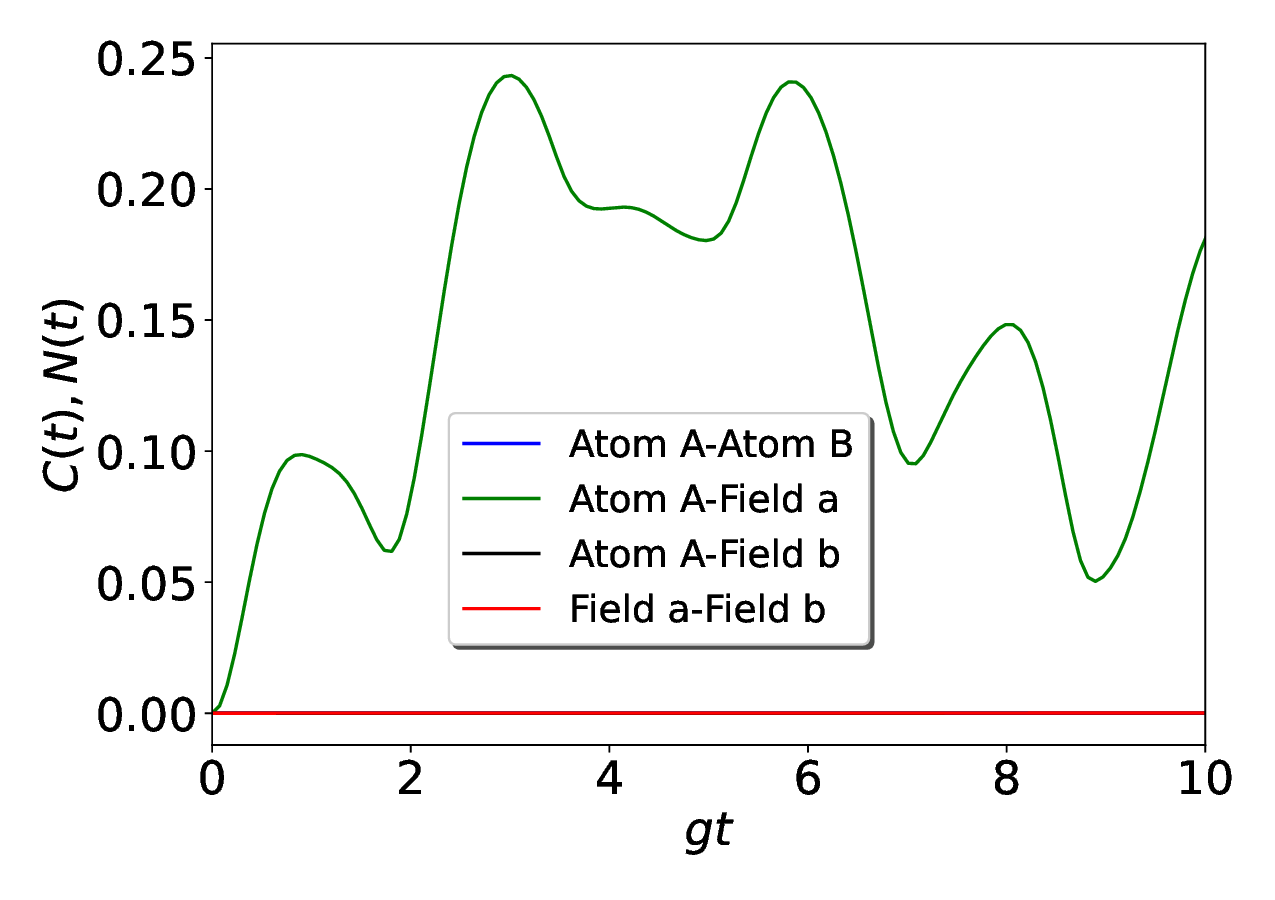}
    \caption*{\textbf{Figure 13.} Entanglement dynamics of atom A-atom B, atom A-field a, atom A-field b and field a-field b subsystems in DJCM for field state SCS  and atomic state in Werner state with $\lambda = 0.25$ and without any Ising interaction between the two atoms.}
\end{figure}
\begin{figure}
    \centering
    \includegraphics[scale = 1.0]{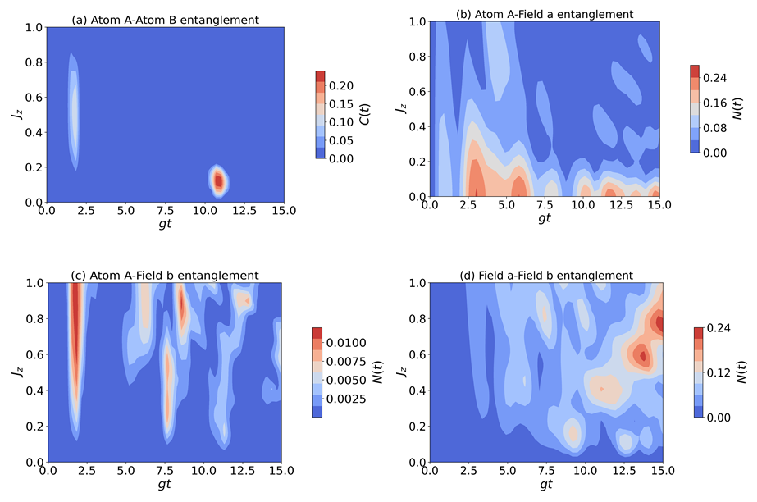}
    \caption*{\textbf{Figure 14.} Contour plots of entanglement dynamics of atom A-atom B, atom A-field a, atom A-field b and field a-field b subsystems with spin-spin Ising interaction in DJCM for SCS and atomic state in Werner ($\lambda = 0.25$). Here, other parameters are $\bar{n}_{c} = 0.5$ and $\bar{n}_{s} = 0.1$. }
\end{figure}
 
\subsubsection{For Werner state}
In the discussion below, the entanglement dynamics of the interacting DJCM when the initial atom A-atom B system is a Werner type state is investigated.  It is well known that the Werner type state is entangled when $\lambda \geq 1/3$  and so we consider two values of $\lambda$ for investigation, $\lambda = 0.75$ and $\lambda = 0.25$. When $\lambda = 0.75$, the initial state is entangled and for $\lambda = 0.25$, it is separable and so studying these two values gives us a comparison between initially entangled state and initially separable state. 

\noindent\underline{{\it For $\lambda = 0.75$}:} \\
The results corresponding to the dynamics of Werner type mixed states is given in Fig. 12 for the DJCM with Ising type interacting cavities. The four possible bipartite entanglements {\it viz} atom A-atom B, atom A-field a, atom A-field b and field a-field b are displayed in Fig. 12. Of the four possible bipartite entanglements, the ESD and revival is present only in the atom A-atom B entanglement. The ESD durations (i.e., the time between the sudden death and revival) get reduced in this model compared to the non-interacting model. Thus the interaction helps in the preservation of entanglement between the atoms. Further, there is no ESD in the atom A-field a bipartite system, the atom A-field b subsystem and the field a-field b subsystems.  

\noindent\underline{{\it For $\lambda = 0.25$}: }\\
In Fig 13, the entanglement dynamics is plotted corresponding to the initial state which is separable.  Of the four possible bipartite systems, only the atom A-field a exhibits entanglement, whereas the remaining three systems do not show any entanglement. The squeezing in the field a causes it to get entangled with atom A, since they are directly interacting. The atom A and field b do not interact directly and so there is no creation of entanglement and the same goes for field a-field b bipartitions. A description of the variation of entanglement with $J_{z}$ and time $gt$ is given in Fig. 14. While the atom A-atom B, atom A-field a and field a-field b entanglement have rich features, the atom A-field b entanglement is present only in negligible amount. 

From the above analysis, it is observed that the Ising interaction creates and also transfers entanglement in the atom-field system. It also makes an initially unentangled system entangled which is evident from Fig. 14.

\begin{figure}
\centering
    \includegraphics[scale = 0.4]{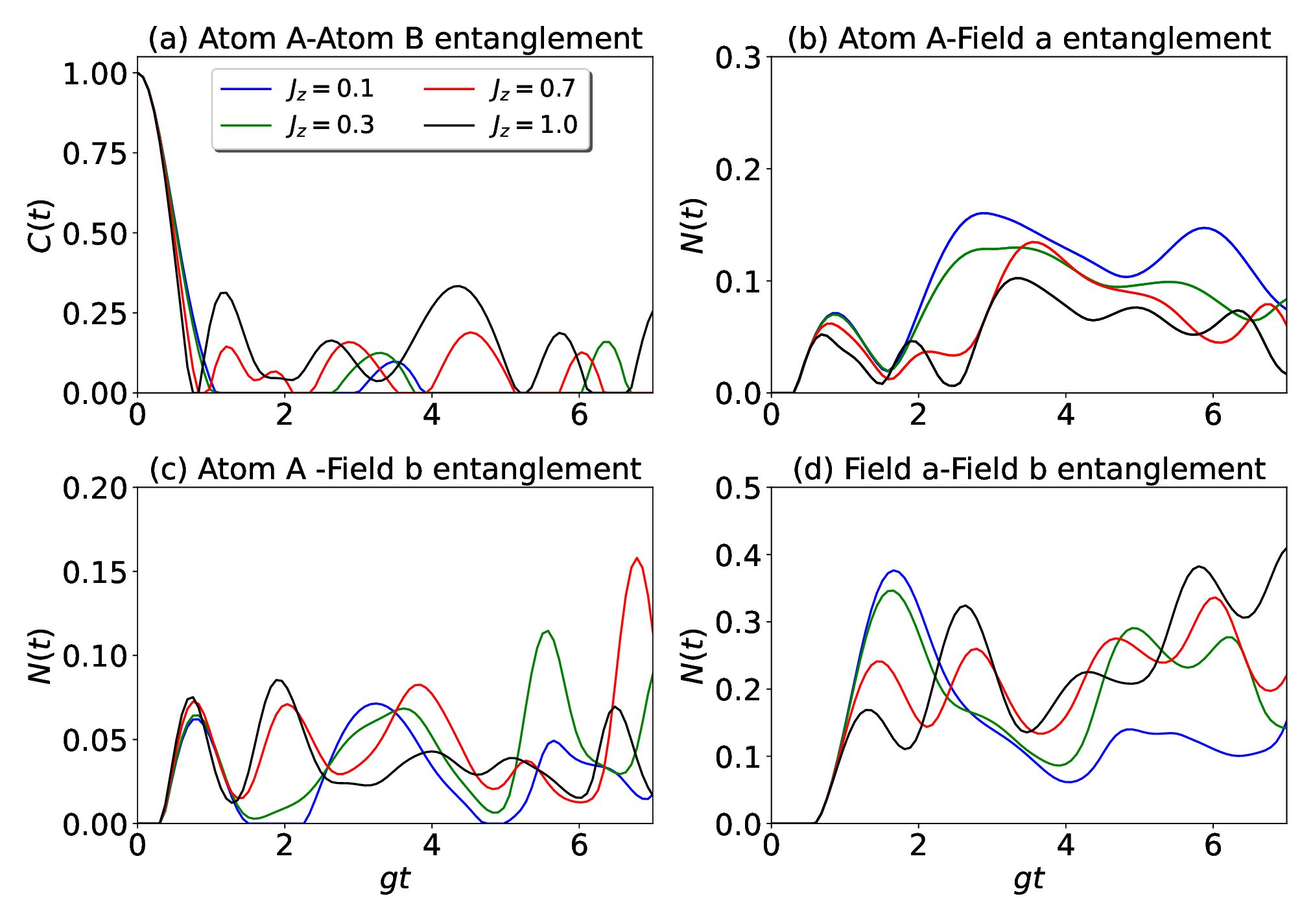}
    \caption*{\textbf{Figure 15.} Effects of spin-spin Ising interaction on entanglement dynamics of atom A-atom B, atom A-field a, atom A-field b and field a-field b subsystems in DJCM for atomic states in the Bell state $\ket{\psi}_\textsubscript{AB}$ and field state in G-L state. The other parameters are $\bar{n}_{c} = 0.5,\bar{n}_{th} = 0.1$, $\theta = \frac{\pi}{4}$ and $J_{z} = 0.1, 0.3, 0.7, 1.0$.}
\end{figure}

\subsection{Glauber-Lachs state}
\subsubsection{For Bell state}
For the Glauber-Lachs (Glauber-Lachs) state, we present the entanglement dynamics in Fig. 15 for a double JC model with Ising interaction. Here, we consider $\bar{n}_c = 0.5$ and $\bar{n}_{th} = 0.1$ and plot for different values of $J_{z}$. The four different bipartite entanglements are given as subplots Fig. 15(a)-(d). From Fig. 15(a) we can see that for $J_{z} = 0.1$, the ESD occurs.  As $J_{z}$ increases, more peaks start appearing reducing the time between the ESD and entanglement revival. For $J_{z} = 1.0$, the ESD phenomenon completely disappears. Comparing with the results corresponding the squeezed coherent states, we observe that the Ising type interaction is more effective in Glauber-Lachs states in removing the ESD. For the atom A-field a entanglement dynamics we find that increasing $J_{z}$ decreases the amplitude of the entanglement in the system as can be seen in Fig. 15(b). In the atom A-field b entanglement, the sudden death feature is present when $J_{z} = 0.1$. When the interaction strength increases the sudden death feature disappears. In the field a-field b entanglement, the initial entanglement is zero and then it increases; also, there is no ESD feature in the dynamics.  

\begin{figure}
    \centering
    \includegraphics[scale = 0.4]{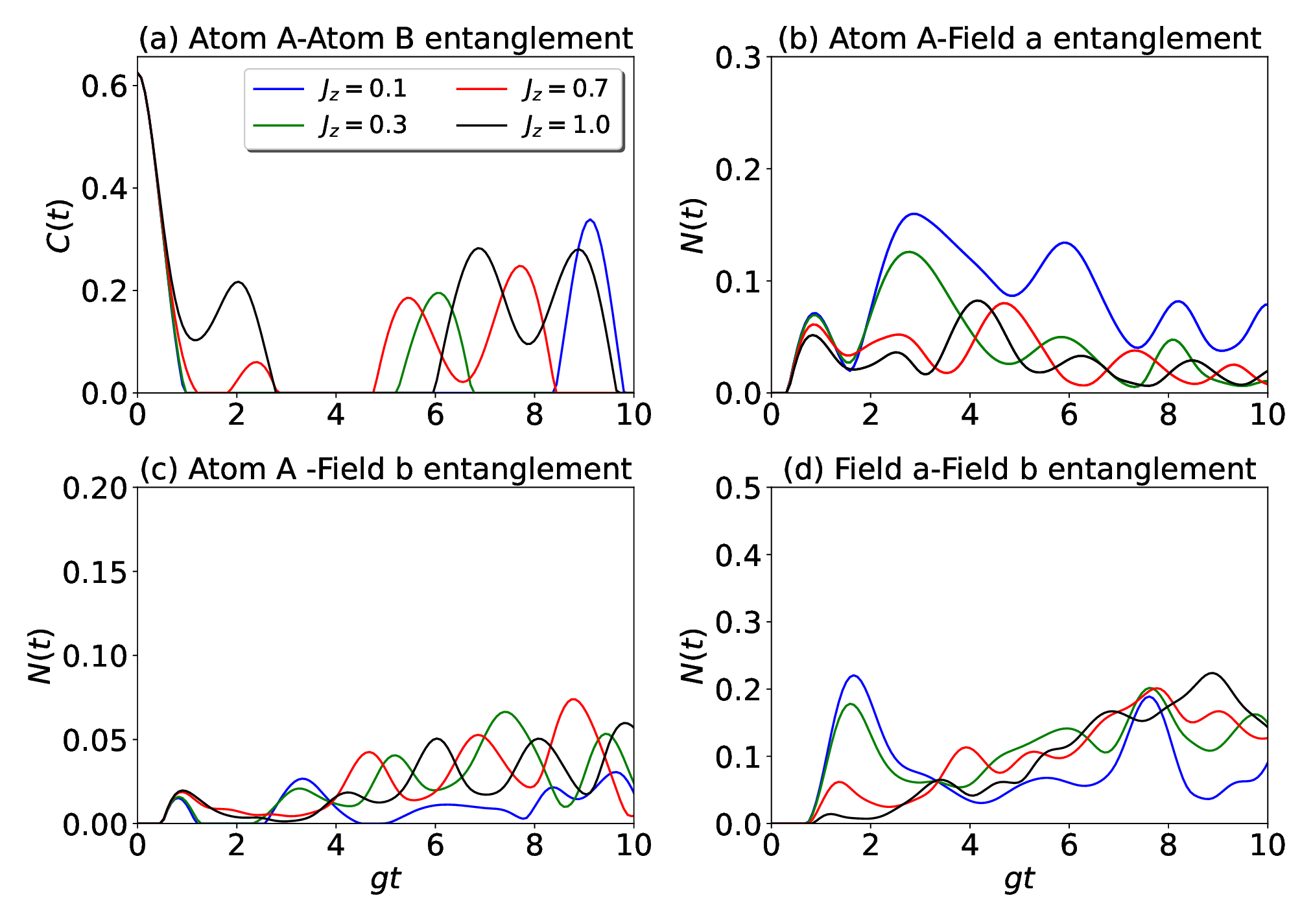}
    \caption*{\textbf{Figure 16.} Effects of spin-spin Ising interaction on entanglement dynamics of atom A-atom B, atom A-field a, atom A-field b and field a-field b subsystems in DJCM with atomic states in Werner state and field state in G-L states. Other parameters are $\bar{n}_{c} = 0.5,\bar{n}_{th} = 0.1$, $\lambda = 0.75$ and $J_{z} = 0.1, 0.3, 0.7, 1.0$.}
\end{figure}

\begin{figure}
    \centering
    \includegraphics[scale = 1]{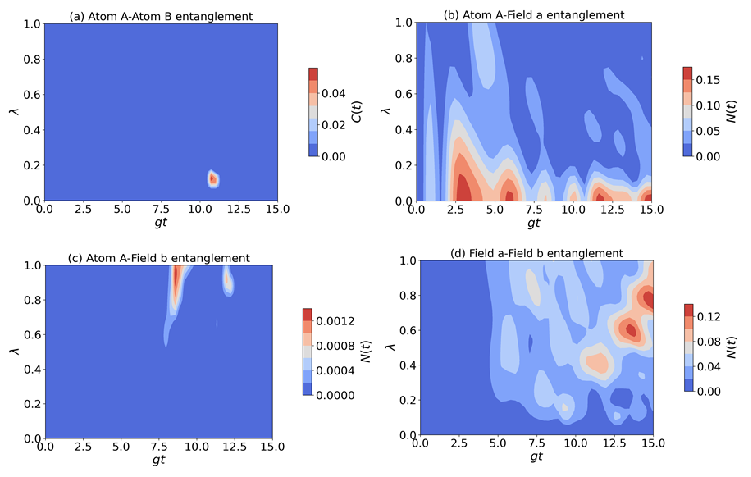}
    \caption*{\textbf{Figure 17.} Contour plots of entanglement dynamics of atom A-atom B, atom A-field a, atom A-field b and field a-field b subsystems with spin-spin Ising interaction in DJCM for Glauber-Lachs states and atomic state in Werner state ($\lambda = 0.25$). Here, other parameters are $\bar{n}_{c} = 0.5$ and $\bar{n}_{th} = 0.1$.}
\end{figure}

\subsubsection{For Werner state}
The dynamics of Werner type state is investigated for two different mixing parameters namely $\lambda =0.75$ as well as $\lambda = 0.25$ and is given in the discussion below.  As stated earlier, the Werner state is entangled for $\lambda = 0.75$ and separable for $\lambda =0.25$ and so their dynamics have quite different characteristic behaviors. We describe below, the dynamics for both these values of $\lambda$.

\noindent\underline{{\it For $\lambda = 0.75$:}}\\
The entanglement dynamics is given in Fig. 16. The four different bipartite entanglements are described in the subplots (a)-(d). The atom A-atom B entanglement and the atom A-field b entanglement show ESDs whereas the atom A-field a and the field a-field b entanglements do not exhibit the behavior. This shows that the interaction preserves the entanglement in bipartite systems.  

\noindent\underline{{\it For $\lambda = 0.25$:}}\\
A description of the entanglement dynamics of the model for the Werner mixing parameter of $\lambda = 0.25$ is shown as a contour plot with the interaction parameter $J_{z}$ and the time $gt$ in Fig. 17. The atom-atom entanglement is almost zero in the system, whereas the atom-field and field-field entanglements are present. The absence of entanglement in the atom-atom subsystem means, the thermal photons try to wash away this entanglement. So, for the Glauber-Lachs states also, the addition of Ising interaction transfers entanglement and it creates entanglement too, however, it is less than as it is observed with SCS.

\begin{figure}
    \centering
    \includegraphics[scale = 0.4]{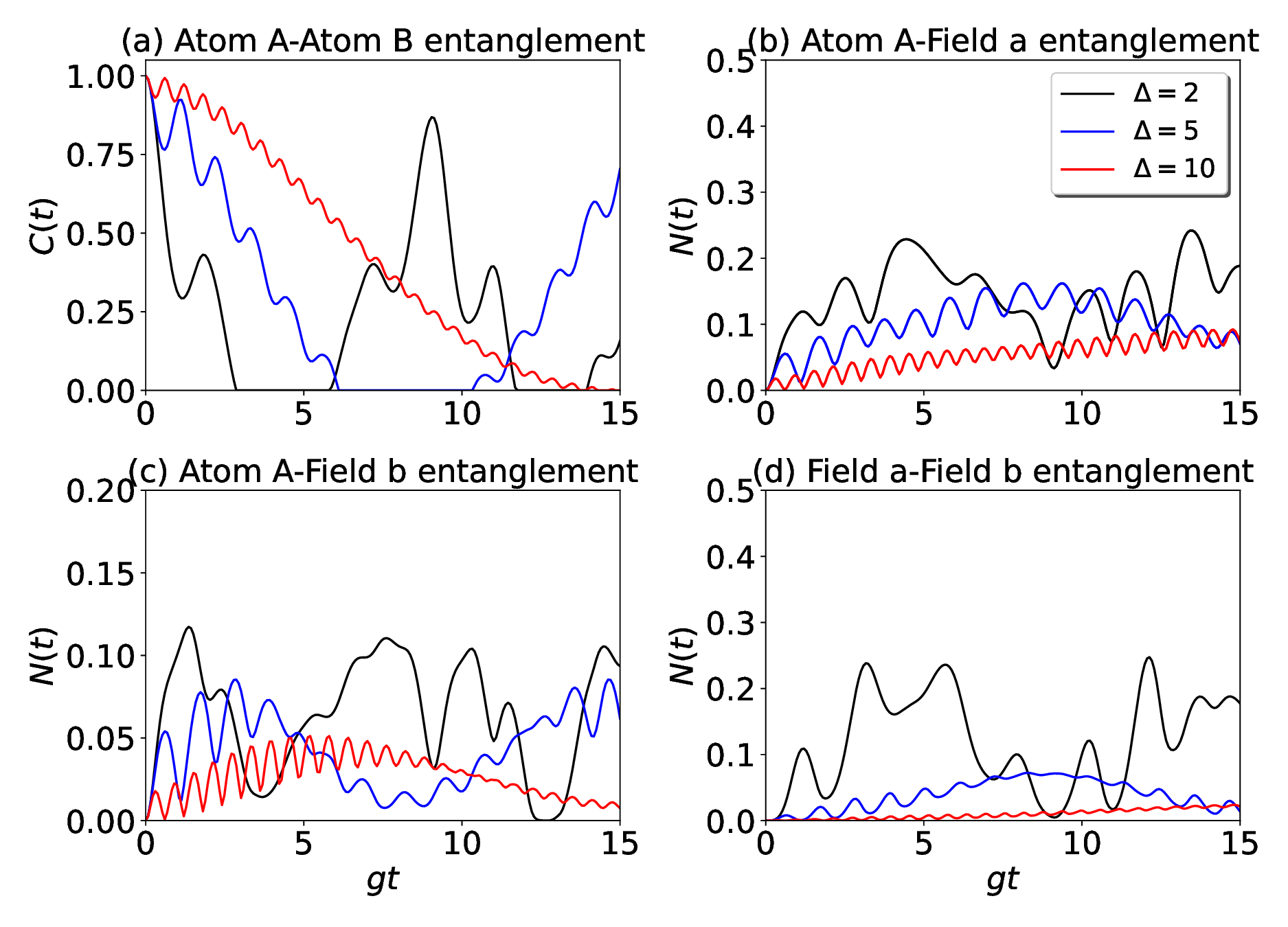}
    \caption*{\textbf{Figure 18.} Effects of detuning on entanglement dynamics of atom atom A-atom B, atom A-field a, atom A-field b and field a-field b subsystems with atomic states in the Bell state $\ket{\psi}_\textsubscript{AB}$ and field state in SCS for $\bar{n}_{c} = 0.5,\bar{n}_{s} = 0.1$, $\theta = \frac{\pi}{4}$ and $\Delta = 2, 5, 10$.}
\end{figure}

\begin{figure}
    \centering
    \includegraphics[scale = 0.4]{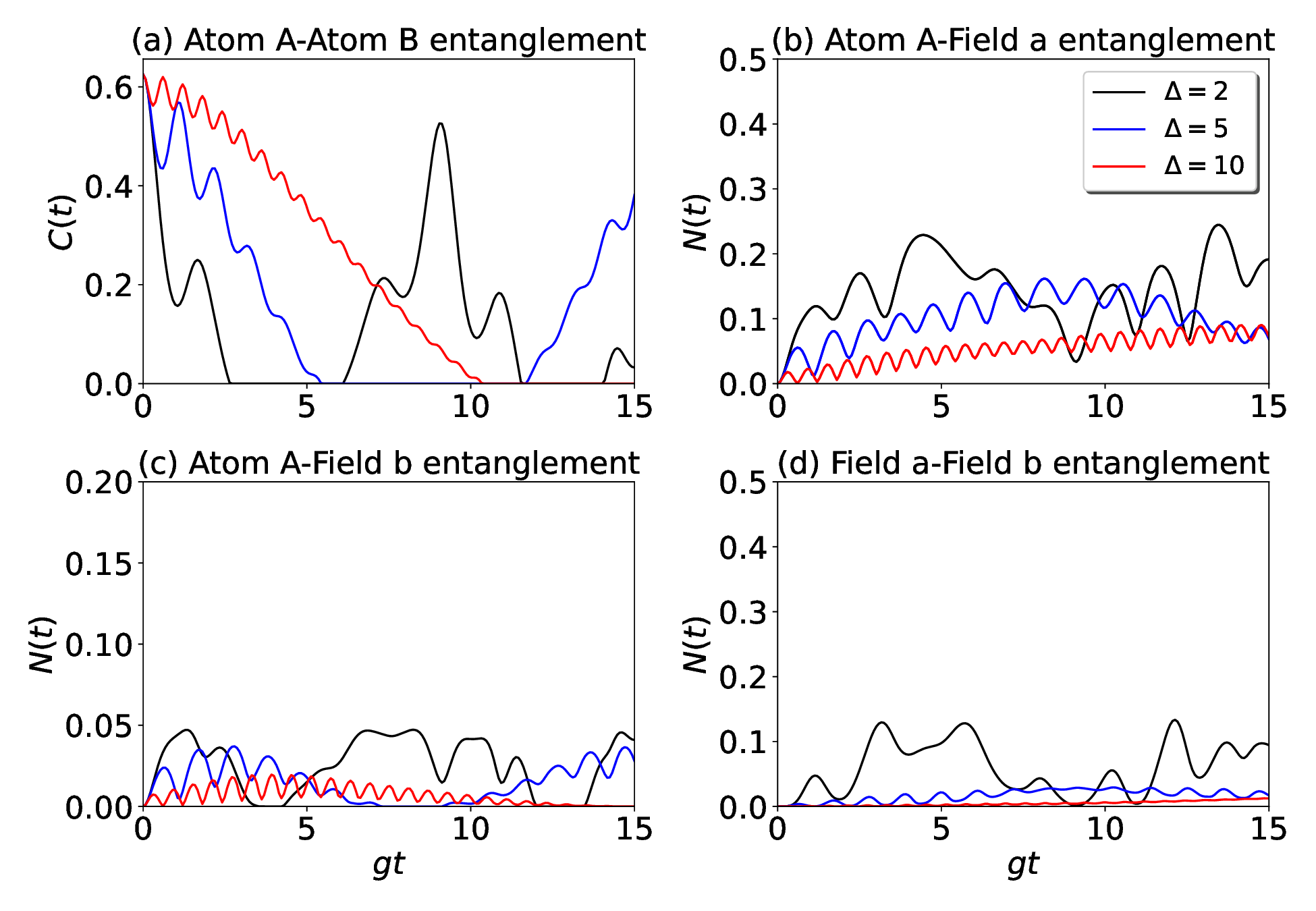}
    \caption*{\textbf{Figure 19.} Effects of detuning on entanglement dynamics of atom A-atom B, atom A-field a, atom A-field b and field a-field b subsystems with atomic states in Werner state and field state in SCS for $\bar{n}_{c} = 0.5,\bar{n}_{s} = 0.1$, $\lambda = 0.75$ and $\Delta = 2, 5, 10$.}
\end{figure}

\section{Effects of detuning on entanglement dynamics}
In the double Jaynes-Cummings Hamiltonian under study, we consider that the frequency of the atomic system is equal to the frequency of the photonic modes. This is not always the case and in some situations these two frequencies are very close to each other, but they are not exactly the same. The difference between the two frequencies is called the detuning and the effective Hamiltonian \cite{lambropoulos2007fundamentals} in this situation becomes
\begin{equation}
\hat{H'}_{\text{eff}} = \Delta \hat{\sigma}_{-}^{\text{A}} \hat{\sigma}_{+}^{\text{A}} +  g\left(\hat{a}^{\dagger} \sigma_{-}^{\text{A}} + \hat{a} \hat{\sigma}_{+}^{\text{A}}\right) + \Delta \hat{\sigma}_{-}^{\text{B}} \hat{\sigma}_{+}^{\text{B}} + g\left(\hat{b}^{\dagger} \hat{\sigma}_{-}^{\text{B}} + \hat{b}\, \hat{\sigma}_{+}^{\text{B}}\right)
\end{equation}
where $\Delta = \omega - \nu$ is the detuning of the atom-field system and is the coupling strength between the atom and the photons. The entanglement dynamics of this mode is discussed for the squeezed coherent states and the Glauber-Lachs state considering both pure and mixed states. 

\subsection{Squeezed Coherent state}
\subsubsection{For Bell state}
The entanglement dynamics of the DJCM with detuning Eq. 15 is given in Fig. 18, where $\bar{n}_{c} = 0.5$, $\bar{n}_{s} = 0.1$ are chosen. The four different bipartite entanglements are given in Fig. 18(a)-(d). Of these it is observed that the ESD is present in the atom A-atom B entanglement given in Fig. 18 (a), albeit the time interval between the sudden death and the revival being reduced. In the rest of the bipartitions {\it viz} the atom A-field a, atom A-field b entanglement as well as the field a-field b entanglement, this feature is no longer present as shown in Fig. 18(b)-(c), and this indicates that the detuning plays a positive role in preserving the entanglement in the system. Further, in all these four types of entanglement, a very highly oscillatory behavior in the dynamics for increasing values of the detuning parameter is observed and this oscillatory behaviour is more prominent for $\Delta = 10$.

\subsubsection{For Werner state}
For the mixed states, we consider a Werner state with the mixing parameter value of $\lambda = 0.75$ and the plots corresponding to this investigation are given in Fig. 19. The results obtained for mixed states are identical to the pure state in the case of atom A-atom B entanglement and the atom A-field a entanglement. The atom A-field b  as well as the field a-field b entanglement values are much low compared to the corresponding values for the pure states.

\begin{figure}
    \centering
    \includegraphics[scale = 0.4]{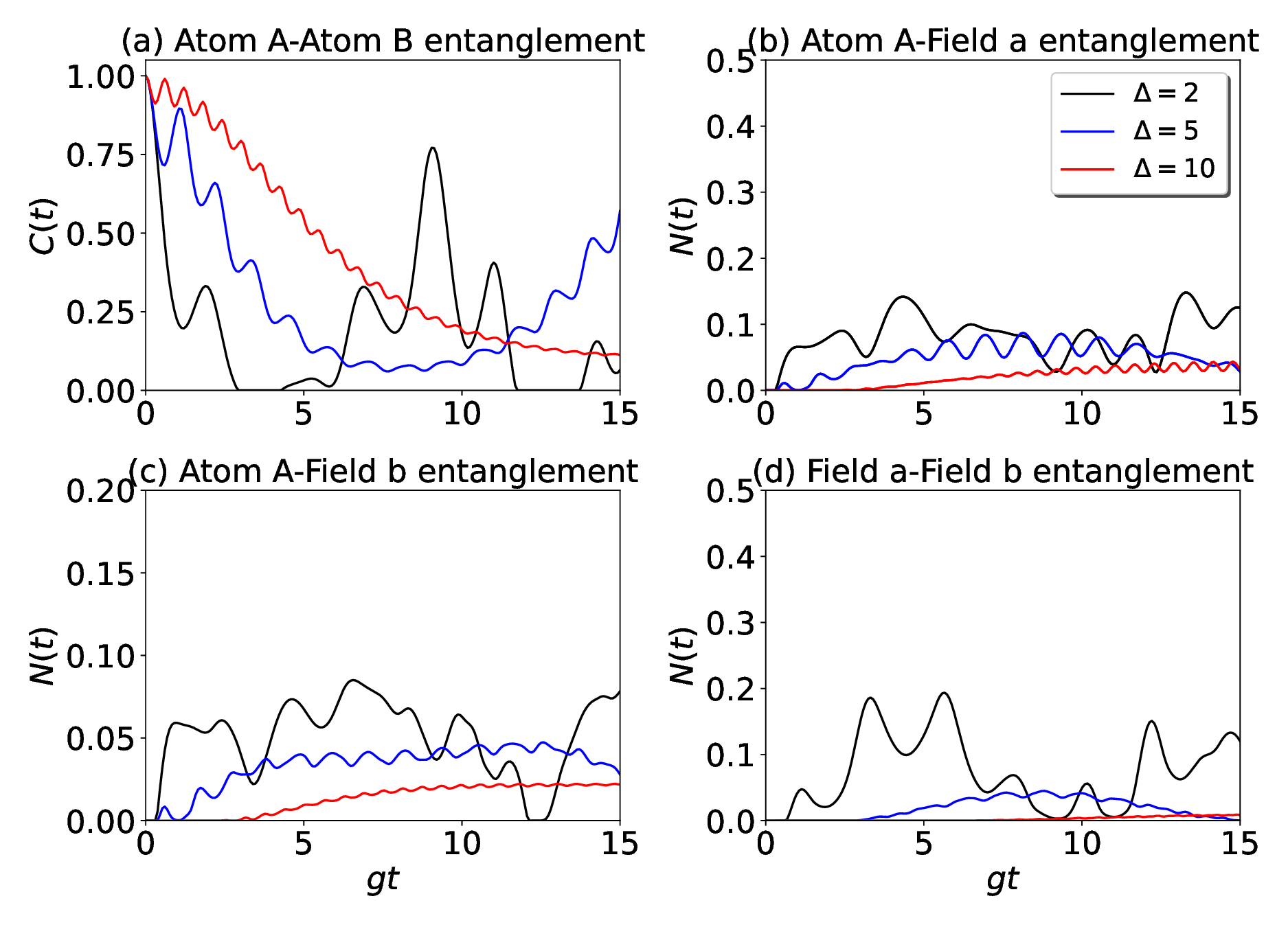}
    \caption*{\textbf{Figure 20.} Effects of detuning on entanglement dynamics of atom A-atom B, atom A-field a, atom A-field b and field a-field b subsystems with atomic states in the Bell state $\ket{\psi}_\textsubscript{AB}$ and field state in Glauber-Lachs states for $\bar{n}_{c} = 0.5,\bar{n}_{th} = 0.1$, $\theta = \frac{\pi}{4}$ and $\Delta = 2, 5, 10$.}
\end{figure}

\begin{figure}
    \centering
    \includegraphics[scale = 0.4]{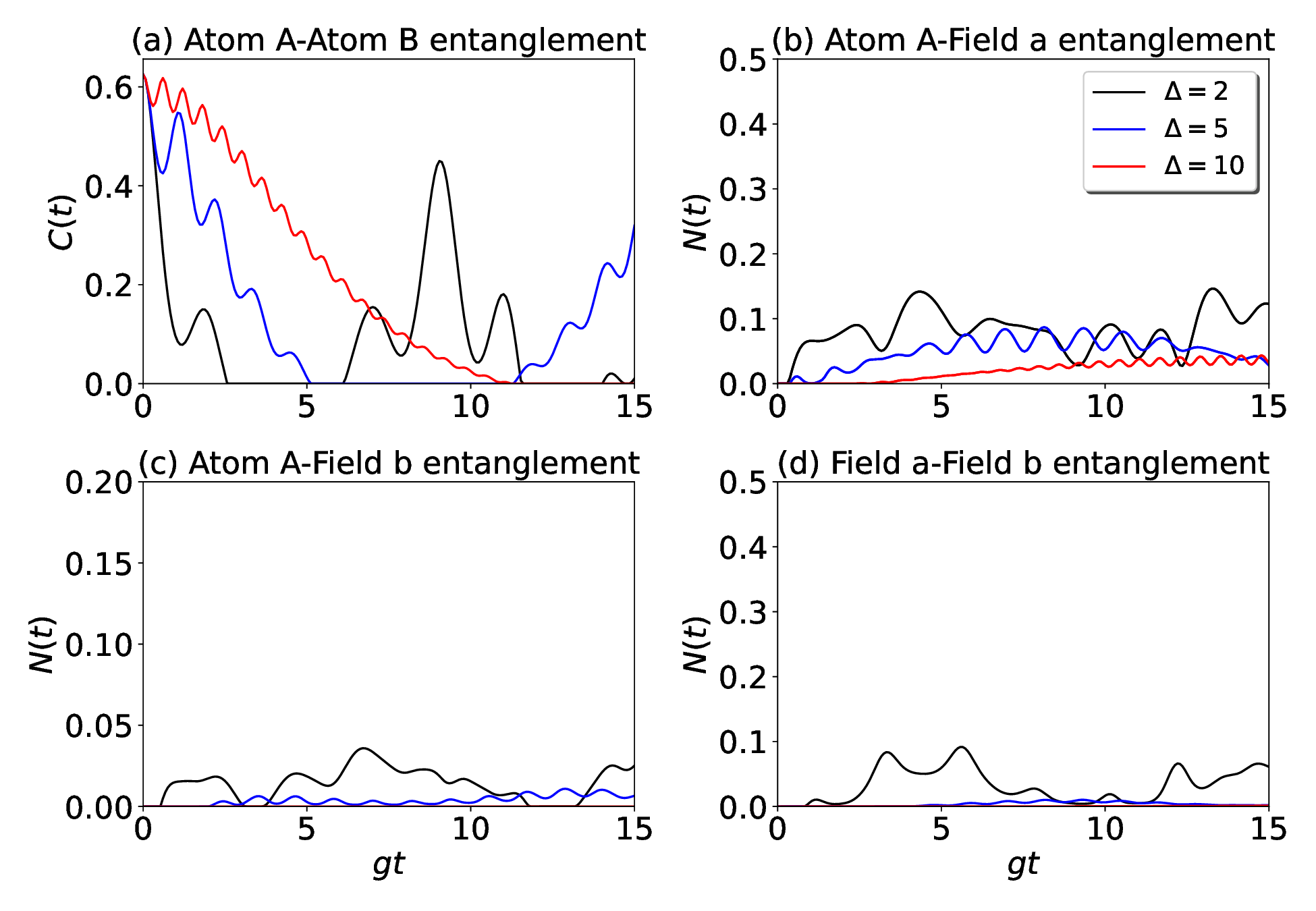}
    \caption*{\textbf{Figure 21.} Effects of detuning on entanglement dynamics of atom A-atom B, atom A-field a, atom A-field b and field a-field b subsystems with atomic states in Werner state and field state in Glauber-Lachs state for $\bar{n}_{c} = 0.5,\bar{n}_{th} = 0.1$, $\lambda = 0.75$ and $\Delta = 2, 5, 10$.}
\end{figure}

\subsection{Glauber-Lachs state}
\subsubsection{For Bell state}
The dynamics corresponding to the Glauber-Lachs state is given in Fig. 20, where the four subplots (a)-(d) give the atom A-atom B entanglement, atom A-field a entanglement, atom A-field b entanglement and the field a-field b entanglements respectively. The atom-atom entanglement displays ESD for small values of detuning ($\Delta=2$). As the value of detuning is increased ($\Delta=5$), the ESD disappears. In the case of atom A-field a entanglement, there is no ESD and the entanglement values decrease with increase in the value of the detuning parameter. This is also observed in the atom A-field b entanglement, but ESD is also present for small values of detuning. Finally, we find that the field a-field b entanglement is high for small values of detuning and decreases with increase in detuning.  

\subsubsection{For Werner state}
In Fig. 21, we show the entanglement dynamics corresponding to the Werner state with mixing parameter of $\lambda = 0.75$ interacting with a Glauber-Lachs state.  Again, we find that the atom-atom entanglement is very pronounced  and displays ESD as well as revival features.  Meanwhile, the atom A-field a entanglement and the field a-field b entanglement does not show any sudden death or revival features. The atom A-field b entanglement is very small and also has the sudden death and revival features. On the whole, we find that the detuning helps in avoiding the ESD. 

For both the radiation states and for both the atomic states, the detuning tries to remove ESD from the entanglement dynamics of all the subsystems, for small value of $\Delta$. However, for higher values of $ \Delta$, the amplitude of atom-atom entanglement  increases very significantly and ESDs are also removed; however, entanglements from other subsystems are decreased and gets transferred to the atom-atom subsystem. So, it can be concluded that detuning is very effective in sustaining atom-atom entanglement in the atom-field system.

\begin{figure}
\centering
    \includegraphics[scale = 0.37]{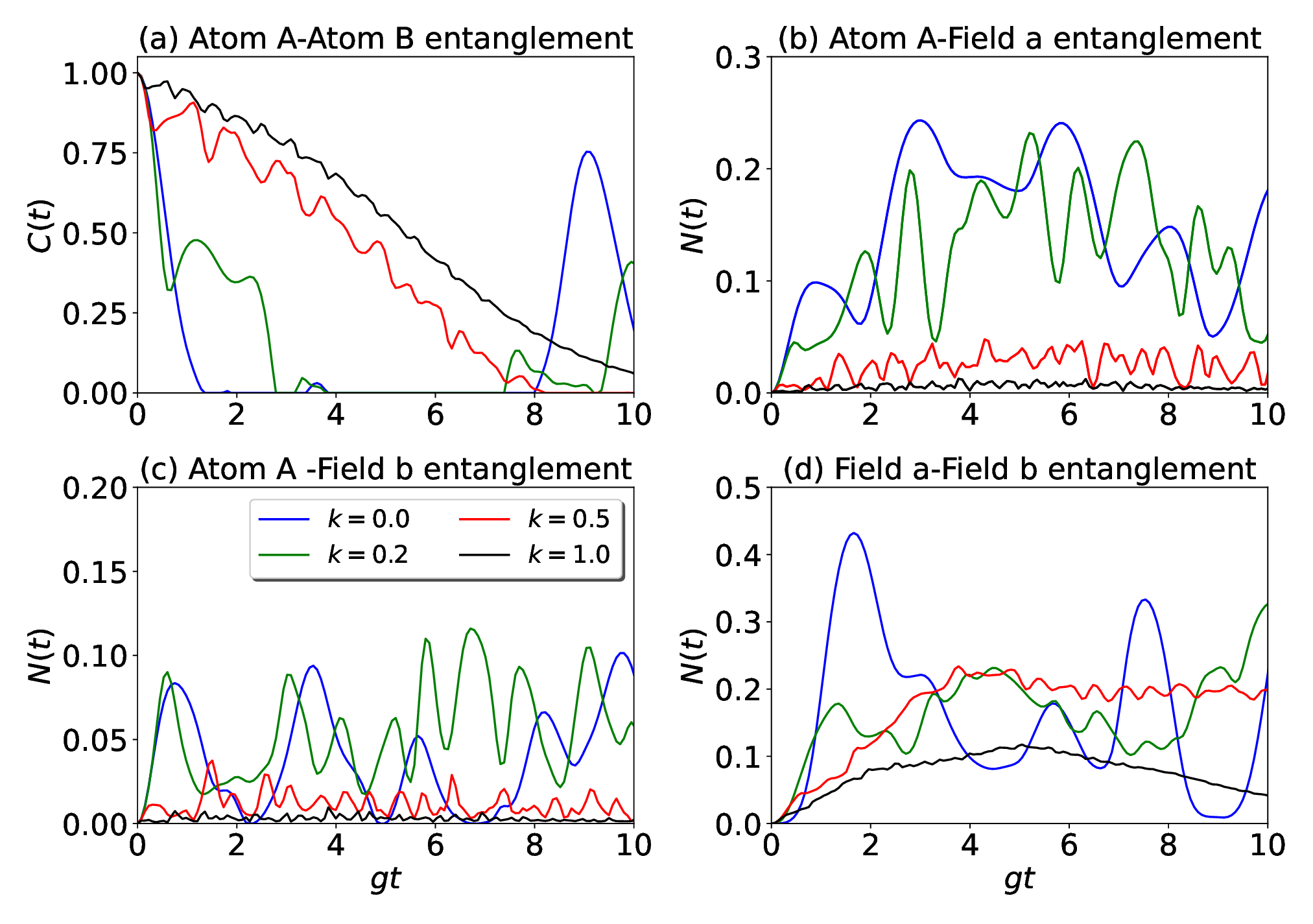}
    \caption*{\textbf{Figure 22.} Effects of Kerr-nonlinearity on entanglement dynamics of atom A-atom B, atom A-field a, atom A-field b and field a-field b subsystems for atomic states in Bell state $\ket{\psi_\textsubscript{AB}}$ and field state in SCS for $\bar{n}_{c} = 0.5,\bar{n}_{s} = 0.1$, $\theta = \frac{\pi}{4}$ and $k = 0.0, 0.2, 0.5, 1.0$.}
\end{figure}

\begin{figure}[ht!]
    \centering
    \includegraphics[scale = 0.4]{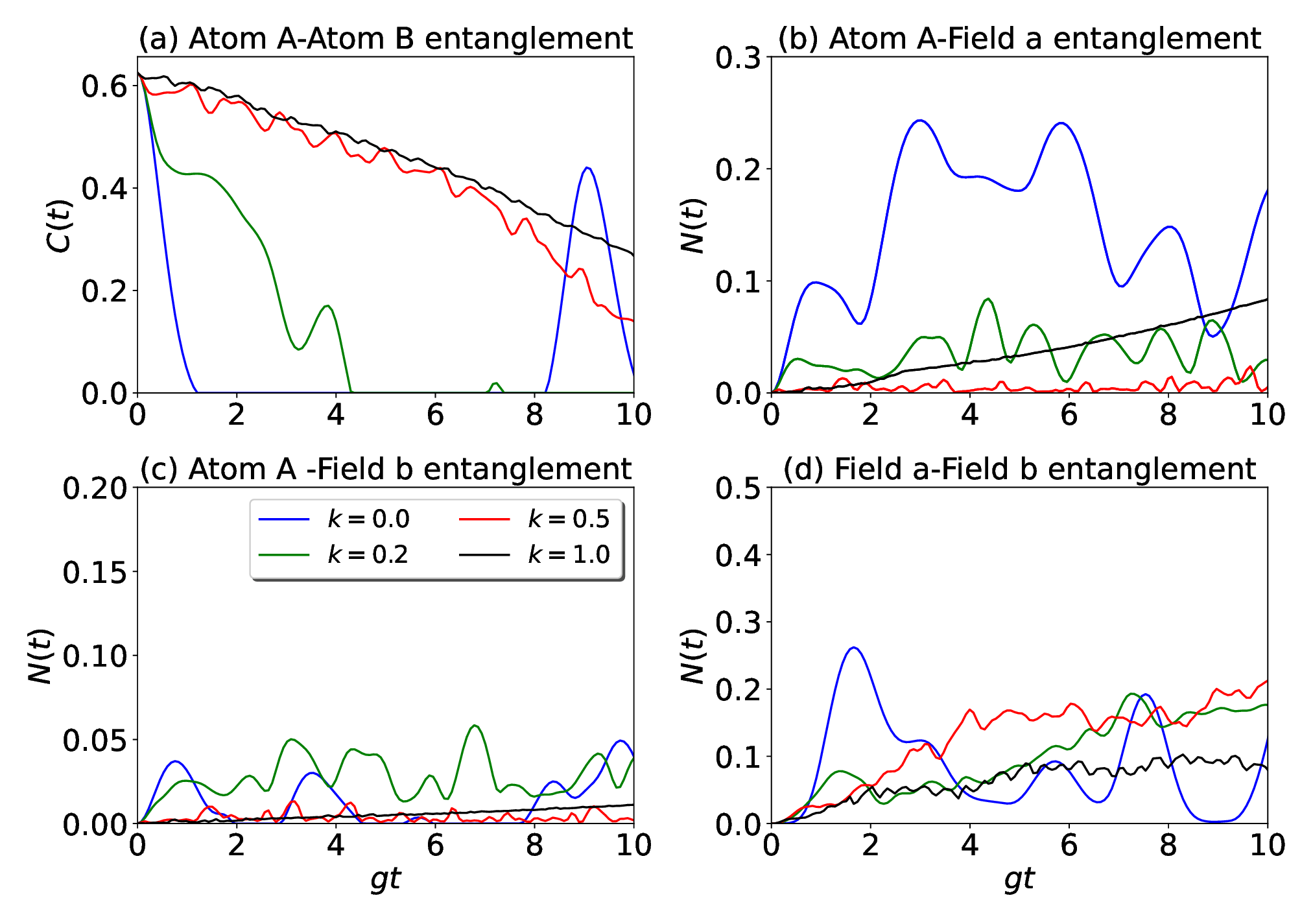}
    \caption*{\textbf{Figure 23.} Effects of Kerr-nonlinearity on entanglement dynamics of atom A-atom B, atom A-field a, atom A-field b and field a-field b subsystems with atomic states in Werner state and field in SCS for $\bar{n}_{c} = 0.5,\bar{n}_{s} = 0.1$, $\lambda = 0.75$ and $k = 0.0, 0.2, 0.5, 1.0$.}
\end{figure}

\section{Effect of Kerr-nonlinearity on the entanglement dynamics}
The DJCM investigated so far has been studied have a linear response.  In order to generalize this to systems with nonlinear response we include a Kerr nonlinear term in our Hamiltonian. The Hamiltonian of a DJCM with a Kerr-nonlinearity is 
\cite{PhysRevA.45.6816, PhysRevA.45.5056, PhysRevA.44.4623, ahmed2009dynamics, sivakumar2004nonlinear, Mo_2022, PhysRevB.105.245310, baghshahi2014entanglement, zheng2017intrinsic, PhysRevA.93.023844}
\begin{align}
\hat{H}_{\text{tot}} =\, \omega \hat{\sigma}_{z}^{\text{A}} + \omega \hat{\sigma}_{z}^{\text{B}} + g \left(\hat{a}^{\dagger} \hat{\sigma}_{-}^{\text{A}} + \hat{a}\, \hat{\sigma}_{+}^{\text{A}}\right) 
+ g \left(\hat{b}^{\dagger} \hat{\sigma}_{-}^{\text{B}} + \hat{b}\, \hat{\sigma}_{+}^{\text{B}}\right) +\,  \nu \hat{a}^{\dagger} \hat{a} + \nu \hat{b}^{\dagger} \hat{b} + \chi \hat{a}^{\dagger 2} {\hat{a}}^{2} + 
\chi \hat{b}^{\dagger 2} \hat{b}^{2},
\end{align}
where $\chi = k \omega$ is the nonlinear coupling constant and $k$ is a non-negative number. The effects of Kerr nonlinearity on the dynamics of atom-field system has been studied in recent years and also in the past. In ref.\cite{mojaveri2018thermal}, the authors have studied the effects of Kerr-nonlinearity and atom-atom coupling on the degree of atom-atom entanglement. Thabet \textit{et. al,} have investigated the effects of Kerr-nonlinearity on the mean photon number, Mandel's $Q$ parameter, entropy squeezing and entanglement dynamics using nonlinear Jaynes-Cummings model\cite{thabet2019dynamics}.
\subsection{Squeezed coherent states}
\subsubsection{For Bell state}
The transient dynamics of entanglement of the above Hamiltonian with Kerr nonlinearity is given in Fig. 22 for the squeezed coherent state when the initial state is a maximally entangled Bell state. Here the subplots Fig. 22(a)-22(d) display the time variation of entanglement of the following four bipartite systems: atom A-atom B entanglement, atom A-field a entanglement, atom A-field b entanglement and field a-field b entanglements respectively. For low values of the Kerr nonlinear parameter we observe the occurrence of ESD and revival in the atom A-atom B system (Fig. 22(a)) and for higher values we do not observe these features. The other bipartite blocks, namely, the atom A-field a system, the atom A-field b system and the field a-field b system do not exhibit the ESD and revival.  Further, there is a complementarity between the atom-atom and field-field entanglement, where when one of them decreases, the other increases suggesting that the entanglement dynamically shifts between the different bipartite systems.  

\subsubsection{For Werner state}
In the case of mixed states we consider a Werner type state with a mixing parameter of $\lambda = 0.75$ and the results are shown in Fig. 23. We consider $\bar{n}_{c} = 0.5$, $\bar{n}_{th} = 0.1$, $\theta = \frac{\pi}{4}$ and plot for different values of the nonlinear parameter. The plots Fig. 23(a)-(d) depict the the atom A-atom B entanglement, atom A-field a entanglement, atom A-field b entanglement and field a-field b entanglements respectively. In the case of atom A-atom B entanglement, the ESD and revival is present, whereas it is not present in the other three bipartite systems. In fact the atom A-field b entanglement is very low compared to the other entanglements.  Also, the atom-atom and field-field entanglements exhibit complementarity in their dynamics. In general, we observe that except for the atom-atom entanglement, the other three types of entanglement increases with increase in the nonlinear parameter.  

\begin{figure}
\centering
    \includegraphics[scale = 0.4]{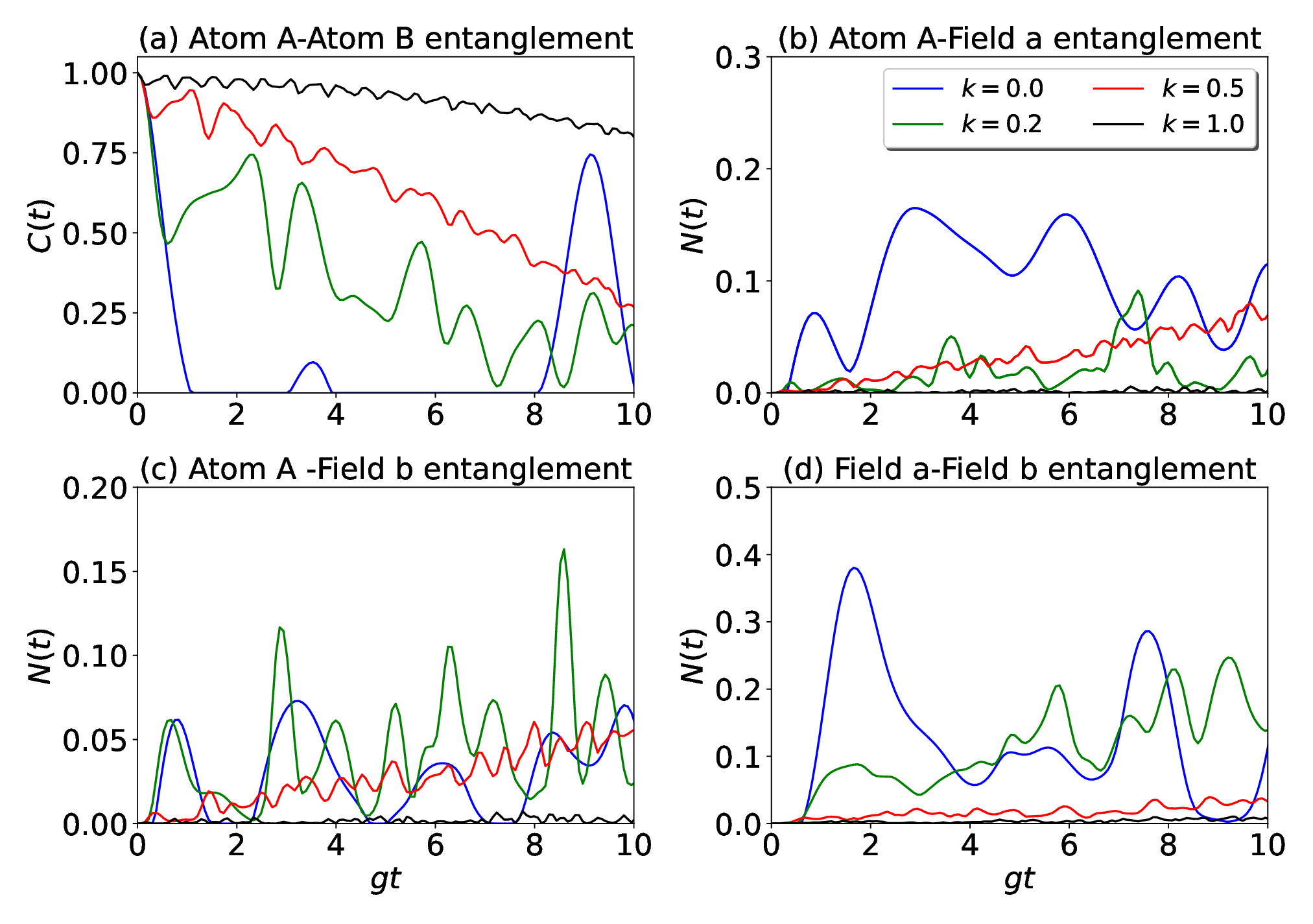}
    \caption*{\textbf{Figure 24.} Effects of Kerr-nonlinearity on entanglement dynamics of atom A-atom B, atom A-field a, atom A-field b and field a-field b subsystems for atomic states in Bell state $\ket{\psi_\textsubscript{AB}}$ and field state in Glauber-Lachs states for $\bar{n}_{c} = 0.5,\bar{n}_{th} = 0.1$, $\theta = \frac{\pi}{4}$ and $k = 0.0, 0.2, 0.5, 1.0$.}
\end{figure}

\begin{figure}[ht!]
    \centering
    \includegraphics[scale = 0.4]{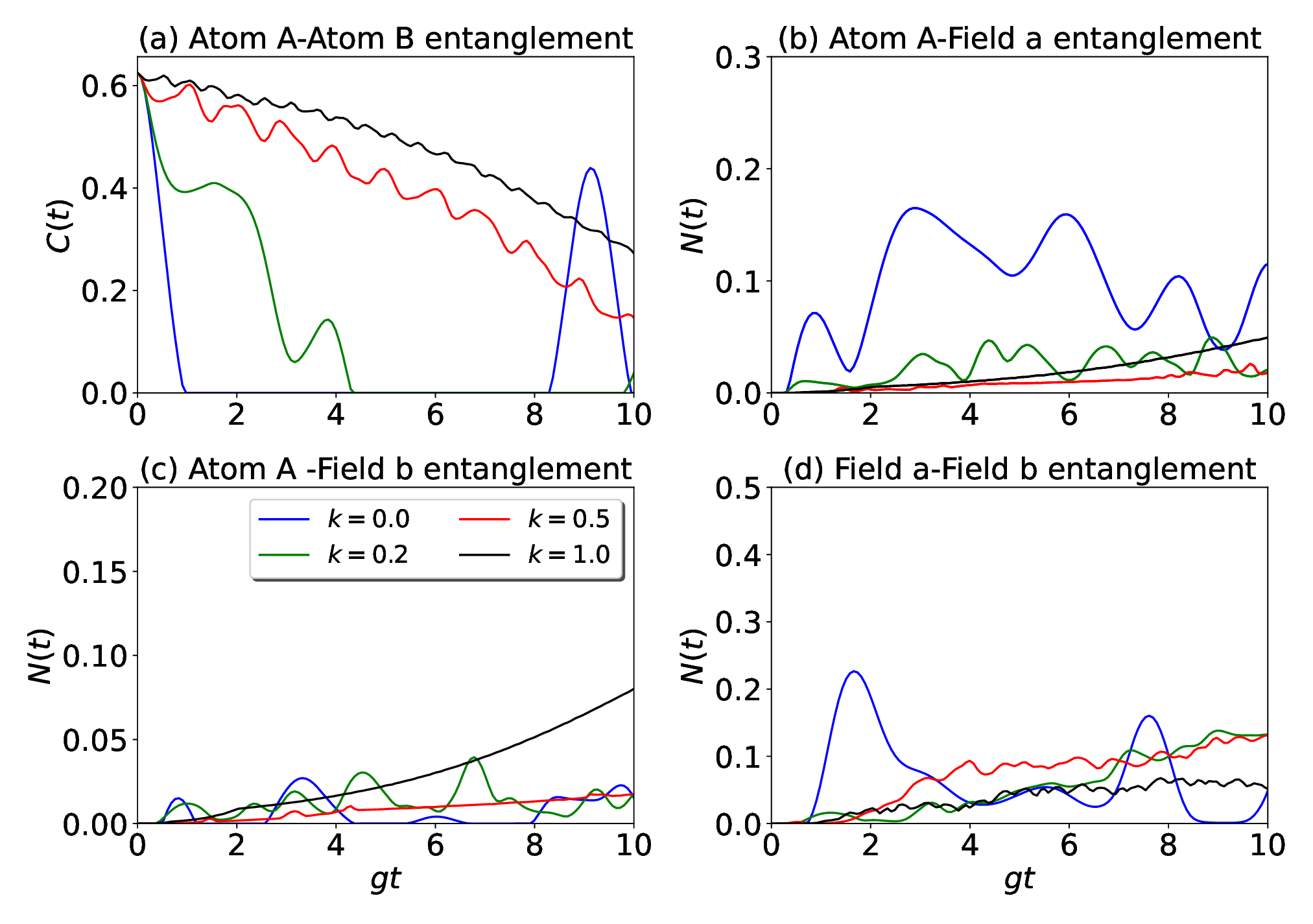}
    \caption*{\textbf{Figure 25.} Effects of Kerr-nonlinearity on entanglement dynamics ofatom A-atom B, atom A-field a, atom A-field b and field a-field b subsystems with atomic states in Werner state and field in Glauber-Lachs states for $\bar{n}_{c} = 0.5,\bar{n}_{th} = 0.1$, $\theta = \frac{\pi}{4}$ and $k = 0.0, 0.2, 0.5, 1.0$.}
\end{figure}

\subsection{Glauber-Lachs state}
\subsubsection{For Bell state}
For initial state which is a maximally entangled Bell state, we give the entanglement dynamics in Fig. 24 for the the radiation mode as a Glauber-Lachs state.  The four subplots Figs. 24(a), 24(b), 24(c) and 24(d) display the entanglement dynamics of the atom A-atom B, atom A-field a, atom A-field b and field a-field b entanglements, respectively. From the dynamics we find that the entanglement falls slowly in the atom A-atom B partition when the Kerr-nonlinearity is high. Further at higher value of Kerr nonlinearity the ESD and revival disappear. For the other three  systems, the entanglement decreases with increase in the Kerr nonlinearity. The atom A-field b system shows ESD and revival.  Also, the field a-field b entanglement shows complementary behavior as observed in the previous discussions.  

\subsubsection{For Werner state}
The variation of entanglement with time for a Werner type mixed state is shown in Fig. 25, for a mixing parameter $\lambda =0.75$. The initial state is entangled and this is observed in the atom A-atom B entanglement in Fig. 25 (a). Further, we also see that the entanglement variation is quite drastic and also shows ESD and revival for lower values of Kerr parameter. Interestingly, the atom A-field a and atom A-field b entanglement show a monotonic increase for $k=1.0$. The field-field entanglement decreases for higher values of the Kerr parameter.

So, we observe that like Ising-type interaction and detuning, Kerr-nonlinearity also effectively transfers the entanglement from other subsystems to the atom-atom subsystem. It helps the atom-atom entanglement to stabilize which is important in many applications.

\section{Conclusion}
This work investigates the entanglement dynamics in a DJCM which is a system of two cavities interacting with two states of radiation. The radiation states in our investigation are chosen as (a) squeezed coherent state, and, (b) Glauber-Lachs state, which is a superposition of thermal and coherent states. We have invesigated the role of various  possible interactions like Ising-type interactions, detuning and Kerr-nonlinearity.  The initial atomic state is prepared in  (a) a maximally entangled Bell state, and,  (b) a mixed state of the Werner-type. Since, the system consists of  two atoms and a radiation mode in each cavity, there are four distinct bipartite entanglements corresponding to (a) atom A-atom B, (b) atom A-field a (c) atom A-field b and (d) field a-field b. The conclusions are briefly mentioned, as follows: 

To start with, we considered each cavity mode to interact with the field inside it. For both Bell and Werner states, it is observed that the effects of $\bar{n}_{th}$ and $\bar{n}_{s}$ on atom A-atom B entanglement are almost similar, viz., the lengths of ESDs increase with the the addition of $n_{s}$ or $n_{th}$.
The atom A-field a and field-field entanglements increase with the increase in $n_{s}$, since squeezing creates entanglement. An important observation is that the squeezing parameter increases the atom-field entanglement, which is not manifest in the atom-atom and atom A-field b entanglements. On the other hand, with increase in the thermal photons, all the four entanglements decrease - which means,  increase of $\bar{n}_{th}$ competes with the coherence in the field and pushes the system to decoherence. The above conclusions are also valid for the Werner-type mixed states.  

In the second part, the Ising interaction among atoms is introduced to study the effects on entanglements. Here, we find that stronger the interaction between the atoms, long-lived is the atom-atom entanglement.  When the atom-atom entanglement is high, the field-field entanglement is low. Thus, we find that the interaction between the atoms preserves the entanglement dynamics in the system. These results are similar for both pure entangled initial states and the Werner-type mixed state.  Next, we investigate the effects of detuning and Kerr-nonlinearity on the entanglement dynamics. 
Detuning also affects entanglements for both the states of radiation field SCS and G-L states with the atoms in pure (Bell) and mixed (Werner) states in a similar way. All the ESDs are removed from the dynamics of the atom-atom subsystem with increasing $\Delta$; whearas, ESDs are observed for other subsystems. So, the increase in $\Delta$ is transferring all the entanglements to the atom-atom subsystem from the other subsystems. Like detuning, proper choice of Kerr-nonlinearity $k$ also removes ESDs from the dynamics of $C(t)$ and $N(t)$. With an increasing value of $k$, $C(t)$ increases, but $N(t)$ decreases creating almost sudden deaths in the entanglement of other subsystems. 
 
This implies that the detuning and Kerr-nonlinearity increase the robustness of entanglement to the external decoherence. It would be interesting to extend this investigation to study a Hamiltonian with more than two cavities, where there are a rich variety of features like entanglement distribution between the cavities.  

\section{Acknowledgements}
The authors thank Professors Arul Lakshminarayan and S. Lakshmi Bala for discussions and suggestions. One of the authors KM is grateful to DST for the financial support from the project PH2021039DSTX008128.

\bibliographystyle{naturemag} 
\bibliography{references_djcm.bib}
\end{document}